\documentclass[aps,prl,amsmath]{revtex4}
\usepackage{graphicx}
\usepackage{color}
\usepackage{bm}
\usepackage{epsfig}
\usepackage{latexsym}
\usepackage{pifont}
\usepackage{float}
\usepackage[utf8]{inputenc}
\graphicspath{{figure/}}
\usepackage{siunitx}
\usepackage{textgreek}

\begin{document}
\newcommand{\be}{\begin{equation}}
\newcommand{\ee}{\end{equation}}
\newcommand{\bea}{\begin{eqnarray}}
\newcommand{\eea}{\end{eqnarray}}
\newcommand{\nn}{\nonumber}

\title{Theoretical model of efficient phagocytosis driven by curved membrane proteins and active cytoskeleton forces}

\author{Raj Kumar Sadhu$^{1}$\footnote{raj-kumar.sadhu@weizmann.ac.il}}
\author{Sarah R Barger$^{2}$}
\author{Samo Peni\v{c}$^{3}$}
\author{Ale\v{s} Igli\v{c}$^{3}$}
\author{Mira Krendel$^{4}$}
\author{Nils C Gauthier$^{5}$}
\author{Nir S Gov$^{1}$\footnote{nir.gov@weizmann.ac.il}}

\affiliation{$^1$Department of Chemical and Biological Physics, Weizmann Institute of Science, Rehovot 7610001, Israel}
\affiliation{$^2$Molecular, Cellular, Developmental Biology, Yale University, New Haven, USA}
\affiliation{$^3$Laboratory of Physics, Faculty of Electrical Engineering, University of Ljubljana, Ljubljana, Slovenia}
\affiliation{$^4$Department of Cell and Developmental Biology, State University of New York Upstate Medical University, Syracuse, USA}
\affiliation{$^5$IFOM, FIRC Institute of Molecular Oncology, Milan, Italy}

\begin{abstract}
Phagocytosis is the process of engulfment and internalization of comparatively large particles by the cell, that plays a central role in the functioning of our immune system. We study the process of phagocytosis by considering a simplified coarse grained model of a three-dimensional vesicle, having uniform adhesion interaction with a rigid particle, in the presence of curved membrane proteins and active cytoskeletal forces. Complete engulfment is achieved when the bending energy cost of the vesicle is balanced by the gain in the adhesion energy. The presence of curved (convex) proteins reduces the bending energy cost by self-organizing with higher density at the highly curved leading edge of the engulfing membrane, which forms the circular rim of the phagocytic cup that wraps around the particle. This allows the engulfment to occur at much smaller adhesion strength. When the curved proteins exert outwards protrusive forces, representing actin polymerization, at the leading edge, we find that engulfment is achieved more quickly and at lower protein density. We consider spherical as well as non-spherical particles, and find that non-spherical particles are more difficult to engulf in comparison to the spherical particles of the same surface area. For non-spherical particles, the engulfment time crucially depends upon the initial orientation of the particles with respect to the vesicle. Our model offers a mechanism for the spontaneous self-organization of the actin cytoskeleton at the phagocytic cup, in good agreement with recent high-resolution experimental observations.
\end{abstract}

\maketitle

\section{Introduction}\label{sen:intro}
Phagocytosis, also termed `cell eating' \cite{Cannon1992,sudha2010,flannagan2012cell,Ellinger2016,Sivakami2021}, is a cellular process by which cells internalize extracellular particles that are relatively large ($0.5-20\mu$m). Phagocytosis plays an important role for the immune system, enabling immune cells to destroy foreign elements and clear dead cells and other debris \cite{Richards_2017}. This process is also important for diagnostic and therapeutic approaches to cancer, for example, using specialized engulfed particles \cite{Imani2017}. Despite its biological importance, it is still not understood how the actin cytoskeleton is coordinated spatio-temporally during the phagocytosis process.

Phagocytosis (and endocytosis) in general requires receptors on the cell membrane that bind to the surface of the target particles \cite{Gao2005}. Binding-driven engulfment, which does not involve the forces of the cytoskeleton, was modelled in \cite{Richards2014,Gao2005,Richards2016,ulrich2019,Jeroen2009,khosravanizadeh2019wrapping}, with purely diffusive dynamics of the receptors \cite{Gao2005,Richards2016} as well as with active drift \cite{Richards2014,Jeroen2009}.  Engulfment of charged particles was previously modelled, and shown to depend on the lateral segregation of mobile charged membrane components (lipids and proteins), leading to a discontinuous wrapping transition \cite{miha2009}.  The effect of signalling molecules was also considered in \cite{Richards2014}, where using drift and diffusion the model successfully recovered the two stages of the process: an initial slow engulfment followed by fast engulfment of the second half of the spherical object, comparing to experiments of neutrophil phagocytosis of polystyrene beads \cite{Richards2014}. Phagocytosis is, however, an active process which is known to involve the forces of actin polymerization that push the engulfing membrane forward \cite{Herant2006}. In subsequent models, detailed description of active forces \cite{Herant2006,Herant2011,Heinrich2015} and also the bending and elastic energies of the membrane and cortex are considered  \cite{bahrami2013,bahrami2014,Serge2015,Liu2006,Tollis2010}. The cytoskeletal components of phagocytosis were investigated during engulfment of artificially produced particles \cite{Jeroen2009,Herant2005,Champion2006,Hendrik2020}, or during bacterial engulfment by the immune cell \cite{Eierhoff2014, Patel2005, Rottner2005}. In \cite{Jeroen2009}, it was found that the engulfment process either stalls before half engulfment of a spherical object, or leads to complete engulfment. In  \cite{Herant2005,Heinrich2005}, the experiment was performed using micro-pipettes to hold both the immune cell and the target particle, allowing the measurement of the cortical tension. Recent experiments provide high-resolution details of the organization of the actin cytoskeleton at the rim of the phagocytic cup \cite{barger2019membrane,Nils2021,OSTROWSKI2019397}, challenging the theoretical modelling of this process.

In reality, however, the shape of the engulfed particles, such as bacteria or viruses, may be highly non-spherical, which motivated several experiments with artificial particles of various geometries \cite{Champion2006,Debjani2013,Gaurav2010,Nishit2010}. In \cite{Champion2006} it was found that the engulfment time for a non-spherical shapes can be five times larger than for a spherical particle \cite{Debjani2013}. In \cite{Gaurav2010} it was observed that an oblate spheroid was engulfed more easily compared to spheres, while spheres were easier to engulf compared to prolate spheroid, rod-shaped particles, or needles \cite{Zhisong2010}. Note that the surface area was not kept constant for the particles with the different geometries. In theoretical studies, several non-spherical shapes were modelled, such as ellipsoids, rod-like particles, and capped cylinder \cite{Dasgupta2013,Dasgupta2014,Robert2011}, however, in these studies the effect of active forces have not been considered.

From these studies we know that the phagocytosis process involves highly complex dynamics of cytoskeleton rearrangement, membrane shape deformations and proteins aggregations  \cite{Richards_2017}. There is at present no theoretical model that explains the dynamics of the self-organization of the membrane and actin cytoskeleton, including the active forces it exerts, during the engulfment process. Here, we address this problem, using a coarse-grained model of a three-dimensional vesicle, having attractive interaction with the target particle. Our model vesicle contains curvature-sensitive proteins, that diffuse on its membrane, and which also recruit actin polymerization, thereby inducing active protrusive forces \cite{miha2019,Sadhu2021}. This modelling approach can be further motivated by the observations of curved membrane proteins that are also related to actin recruitment at the leading edge of the cell membrane \cite{joern2014,begemann2019mechanochemical,pipathsouk2021wave}. Using this model we are able to expose the role of curved membrane proteins, coupled with adhesion and active cytoskeletal forces, during the phagocytosis process.


\section{Results}\label{sec:result}
We simulate the dynamics of a three-dimensional vesicle,  which is described by a closed surface having $N$ vertices, each of them connected to its neighbours with bonds, forming a dynamically triangulated, self-avoiding network, with the topology of a sphere \cite{Sadhu2021,miha2019}. The details of the model are given in the Materials and Methods section (also see SI sec. S1). We first study the case where the phagocytic particle is spherical, and then generalize it for non-spherical particles as well.

\subsection{Spherical particles - Passive curved proteins} \label{sec:spherical_passive}
As a first validation of our model, we simulated the engulfment process of a spherical particle by a protein-free vesicle (see SI secs. S2-S3, Figs.S2-S5; Supplementary Movies-S1,S2). In agreement with previous studies \cite{Lipowsky1998} we find an optimal size for full engulfment of the particle in this system; very small particles can not be engulfed due to the high bending energy, while very large particles require larger vesicles in order to be fully engulfed. An optimal radius of the spherical particle was found to be $R \sim 10~ l_{min}$ (where $l_{min}$ is the minimal allowed length of an edge on the triangulated surface). The protein-free case can be compared to a simplified analytical model (see SI sec. S3), and used to calibrate the relevant parameter regime.  
\begin{figure}[ht]
\centering
\includegraphics[scale=0.9]{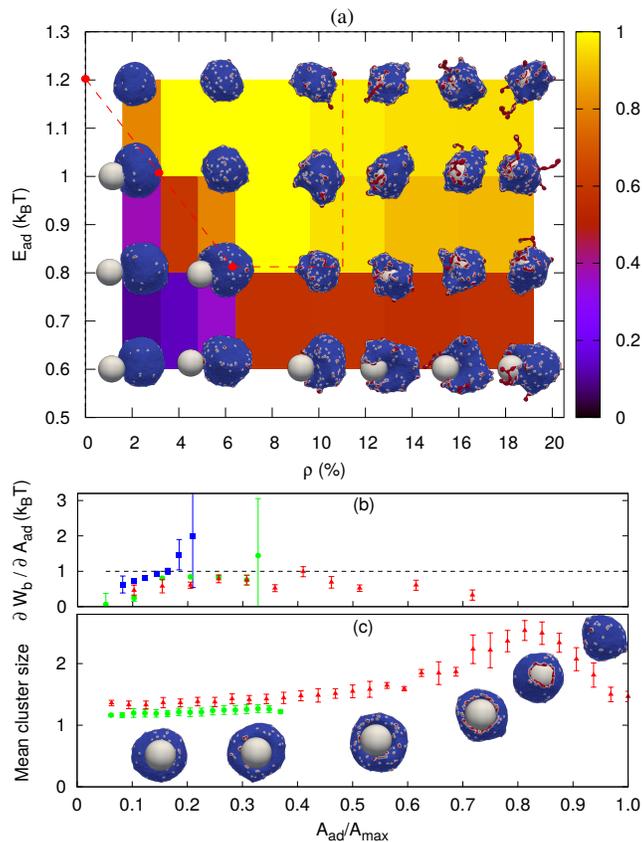}
\caption{Sphere engulfment by a vesicle with passive proteins. (a) Phase diagram in the $E_{ad}-\rho$ plane. The red dashed line is separating the fully engulfed state (above the line) and the partially engulfed state. The snapshots are shown for $E_{ad}=0.60, 0.80, 1.0, 1.20$ (in units of $k_B T$) and $\rho=1.6~\%, 4.8~\%, 9.6~\%, 12.8~\%, 16\%, 19.2~\%$. (b) Comparison of adhesion energy gain (dashed horizontal line) and bending energy cost ($\partial W_b/\partial A_{ad}$, points) per adhered node (both having units of $k_BT$) as function of the engulfment fraction, for a fixed adhesion strength $E_{ad}=1.0~ k_BT$ and various values of $\rho$. Blue symbols denote $\rho=0.8 ~\%$, green circles for $\rho=2.4 ~\%$ and red triangles for $\rho=4.0~ \%$. (c) Mean cluster size (mean number of proteins in a cluster) as function of the engulfed area fraction, for both partial and complete engulfment cases. The colors codes are same as fig. \ref{fig:passive}b. }
\label{fig:passive}
\end{figure}

Next, we consider vesicles with passive ($F=0$, where $F$ is the active protrusive force exerted by the proteins), curved proteins. In Fig. \ref{fig:passive}a, we show the phase diagram in the $\rho-E_{ad}$ plane (where $E_{ad}$ is the adhesion energy per adhered node, and $\rho$ the areal density of the proteins on the vesicle membrane). For small $E_{ad}$, below $\sim 0.60~k_BT$, we only obtain partial engulfment, with the engulfment area increasing with $\rho$ but not reaching full engulfment. For intermediate $E_{ad} (0.80-1.0~k_B T)$, there is an incomplete engulfment for small $\rho$ (Movie-S3), but the engulfment is complete for larger $\rho$ (Movie-S4). This is the interesting regime, where engulfment is facilitated by the presence of the curved proteins. 

Note that we calculate the adhered fraction of the spherical particle ($A_{ad}/A_{max}$), using for $A_{max}$ the maximum adhered area. This may be larger than the true surface area of the particle, since we have a finite width of the adhesive interactions, and within this width the vesicle may fluctuate (for more details see SI, sec. S2).

Increasing $\rho$ reduces the bending energy cost of the engulfment by the spontaneous aggregation of the curved proteins at the sharp leading edge of the engulfing membrane, and allows the vesicle to adhere more and engulf the particle. This is the same mechanism that allows passive curved proteins to drive vesicle spreading on flat adhesive substrates \cite{Sadhu2021}.

The engulfment process can be understood by considering the competition between adhesion and bending energy cost of the vesicle as the engulfment proceeds. In Fig. \ref{fig:passive}b we compare the adhesion energy gain ($E_{ad}$) and bending energy cost ($\partial W_b/\partial A_{ad}$), per additional adhered membrane node, as function of the adhered fraction of the spherical particle. The bending energy cost increases with adhered fraction, and for small $\rho$ the engulfment process stops where the bending energy cost becomes higher than the adhesion energy gain, per node. In Fig. \ref{fig:passive}b we show two cases of partial engulfment, for $\rho=0.8 ~\%,2.4 ~\%$ (blue and green points), where the bending energy cost becomes larger than the adhesion energy gain, and the system reaches steady-state at $A_{ad}/A_{max}\sim 0.2, 0.35$ respectively. At these low densities the curved proteins fail to form large aggregates at the leading-edge of the engulfing membrane, as shown by the mean cluster size (Fig. \ref{fig:passive}c), and therefore do not lower the bending energy cost sufficiently (see insets of Fig. \ref{fig:passive}c for the adhered fractions that are smaller than 0.5).

Complete engulfment is achieved for higher densities, where the adhesion energy gain is larger than the bending energy cost. For $\rho=4.0 ~\%$ the bending energy cost is always smaller than the adhesion energy gain, and the system proceeds to complete engulfment (red points in Fig. \ref{fig:passive}b). This is driven by the formation of a large protein cluster along the leading-edge of the engulfing membrane (Fig. \ref{fig:passive}c and insets for adhered fractions that are larger than 0.5). 

Note that in addition to the adhesion energy gain, there is also the gain in protein-protein binding energy, which further helps drive the membrane over the bending energy barrier, but this is a small contribution due to the small size of the ring-like clusters of proteins (Fig. \ref{fig:passive}c). Following the completion of the engulfment, this ring-like cluster of curved proteins disperses spontaneously. Since the topology of the membrane is fixed, we do not allow for membrane fusion, and there remains a tiny open membrane neck that connects the engulfed sphere with the space outside the vesicle.

For very large $\rho$ (Fig. \ref{fig:passive}a), however, the wrapping is again incomplete. The large number of curved proteins stabilize the cluster at the membrane rim and prevent the hole from closing up (Movie-S5). For large $E_{ad} (\geq 1.2~k_BT)$, the adhesion interaction between vesicle and particle is strong enough to drive complete engulfment even in the absence of curved proteins. The red dotted line in Fig. \ref{fig:passive}a encloses the region within which we find complete engulfment. The shape of this complete-engulfment transition line can be qualitatively reproduced by a simplified analytical model (SI sec. S4; Figs. S7, S8). 

Finally, we also calculated the engulfment time as function of $\rho$ (see SI sec. S5, Fig. S9), where we find that it is decreasing with increasing $\rho$. The engulfment time diverges for $\rho \sim 2.25 ~\%$, below which there is no engulfment (for $E_{ad}=1.0 ~k_BT$).

\subsection{Spherical particles - Active curved proteins} 
\label{sec:spherical_active}

\begin{figure}[!htbp]
\centering
\includegraphics[scale=0.85]{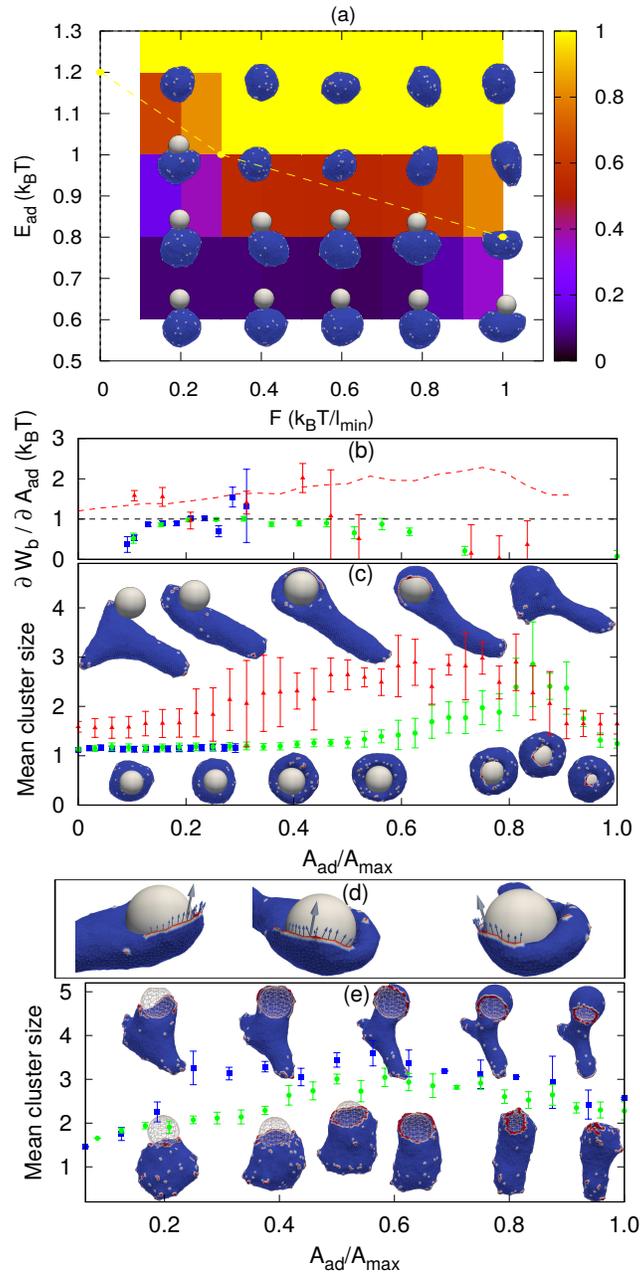}
\caption{Sphere engulfment by a vesicle with active proteins. (a) Phase diagram in the $E_{ad}-F$ plane for a low concentration of curved proteins $\rho=1.6~\%$. The dashed yellow line is the transition from the fully engulfed (above the line) to partially-engulfed states. The snapshots are shown for $E_{ad}=0.60, 0.80, 1.0, 1.20$ (in units of $k_B T$) and $F=0.20, 0.40, 0.60, 0.80, 1.0$ (in units of $k_B T/l_{min}$). (b) Comparison of bending energy cost with the adhesion energy gain (both having units of $k_BT$ per adhered node) for different values of $F$. The blue boxes, green circles and the red triangles are for $F=0.1$, $0.4$ and $2$ (in units of $k_BT/l_{min}$) respectively. The red dashed line is showing the combined effective energy gain per adhered node, including the adhesion and the work done by the active forces (for $F=2~k_BT/l_{min}$). (c) The mean cluster size as function of engulfed area fraction, for the same cases shown in (b) (same color code). Lower snap-shots correspond to $F=0.1,0.4$, and top snap-shots to $F=2$. (d) Snap-shots of the vesicle ($F=2 ~k_BT/l_{min}$) indicating the contribution of the active forces to the the engulfment. Arrows show the tangential component of the individual forces (small arrows), and the direction of the total force (large arrow), for proteins that are close to the spherical surface. For (b-d), we use $E_{ad}=1 ~k_BT$ and $\rho=1.6~\%$. (e) Mean cluster size for large $E_{ad}$ and $\rho$. Green circles are for $\rho=6.4\%$, $F=1.0~k_BT/l_{min}$, and blue boxes are for $\rho=4.8~\%$, $F=2.0~k_BT/l_{min}$. We use $E_{ad}=1.5 ~k_BT$ for both the cases. }
\label{fig:active}
\end{figure}

Next, we study the engulfment by a vesicle containing active curved proteins (Fig. \ref{fig:active}). In order to expose most clearly the role of activity during the engulfment process, we start with a low concentration of curved proteins ($\rho=1.6~\%$), such that their passive effect is weak (Fig. \ref{fig:passive}a). In Fig. \ref{fig:active}a, we show the engulfment phase diagram in the $F-E_{ad}$ plane. We find that the active forces enable full engulfment even at low adhesion strengths, where the passive curved proteins ($F=0$) are not sufficient to drive full engulfment. For small $E_{ad}$, the engulfment is incomplete even for large $F$, while for intermediate $E_{ad}$ the engulfment becomes complete as $F$ increases (Movie-S6,S7,S8). For large values of $E_{ad}$, the engulfment is achieved even for the passive system. 

As for the passive case, we analyze the engulfment process by plotting the adhesion energy gain and the bending energy cost per adhered node, as function of $A_{ad}/A_{max}$. In Fig. \ref{fig:active}b we compare both the case of complete and incomplete engulfment, as the force is increased. As in the passive case, incomplete engulfment occurs when the bending energy cost per node becomes larger than $E_{ad}$ (blue boxes in Fig. \ref{fig:active}b). For a higher force ($F=0.40 ~k_BT/l_{min}$), the active force enhances the aggregation of the proteins around the rim and thereby lowers the bending energy cost, resulting in complete engulfment (green circles in Fig. \ref{fig:active}b). 

Even higher forces ($F=2 ~k_BT/l_{min}$), produce large shape fluctuations of the vesicle, and the bending energy cost increases. However, a complete engulfment is achieved nevertheless (red triangles in Fig. \ref{fig:active}b, and top insets in Fig. \ref{fig:active}c). In this case, the engulfment is enabled by the active forces that directly push the membrane over the spherical surface, and complete engulfment is achieved even when the bending energy cost per node goes above $E_{ad}$. In order to extract the contribution due to active forces in this case, we define a region close to the particle (phagocytic cup or rim) and assume that only the forces from these proteins are helping the vesicle to engulf the particle. We then calculate the magnitude of active forces in the tangential direction to the surface of the particle (Fig. \ref{fig:active}d). We then divide this total force $F_{total}$ by the length of the leading rim of the membrane, to extract the effective active work done per adhered node (see SI sec. S6, Fig. S10 for details). We add this active work to the adhesion energy gain per adhered node (red dashed line in Fig. \ref{fig:active}b), which is shown to be sufficient to offset the bending energy cost per node. 

Note that in the active case the complete engulfment can proceed at much lower densities of proteins, compared to the passive case, such that there are not necessarily enough of them to complete a ring-like cluster at the leading edge. An arc-like cluster of active proteins, that forms at an earlier stage of the engulfment process (red triangles in Fig. \ref{fig:active}c), can be sufficient to drive a lamellipodia-like protrusion that spreads over the spherical particle to complete the engulfment (Fig. \ref{fig:active}(c,d)). This is further shown in Fig. \ref{fig:active}e, for larger protein densities (Movie-S9,S10). The spherical particle is transparent to ensure the visibility of the protein clusters along the rim of the phagocytic cup. Due to the higher density the rim clusters form at lower adhered fraction (compared to Fig. \ref{fig:active}c), but they are still highly fragmented and do not merge to form a complete circular ring until very late stages of the engulfment (Fig. \ref{fig:active}e). 

Many of the spatio-temporal features that we found for the curved active proteins in our simulations match very well the recent high-resolution experimental images of the actin organization along the leading edge of the engulfing membrane protrusion during phagocytosis (Fig. \ref{fig:experiment}, Movie-S11,S12) \cite{Nils2021}. In both the experiments and the simulations we find that the actin ring is highly fragmented, in the form of dispersed "teeth" and "arcs" along the leading edge. The actin aggregate becomes more cohesive in the final stages of the engulfment, before it disperses after the engulfment is complete (Also see SI sec. S7, Movie-S13).

\subsection{Spherical particles - Engulfment dynamics} 
\label{sec:spherical_time}

The engulfment dynamics for the different systems are shown in Fig. \ref{fig:eng-dynamics}a. For the protein-free and passive proteins systems, driven by adhesion alone, we find in our simulations the typical "two-stage" behavior: an initial slow growth up to adhered area fraction of $\sim0.5$, and then a very fast stage to complete engulfment. A similar behavior is also observed for the active system in the regime of low protein densities. This dynamics was observed experimentally, and for the passive engulfment system was also motivated by theoretical modelling \cite{Richards2014,miha2009}. 

\begin{figure}[h]
\centering
\includegraphics[scale=0.47]{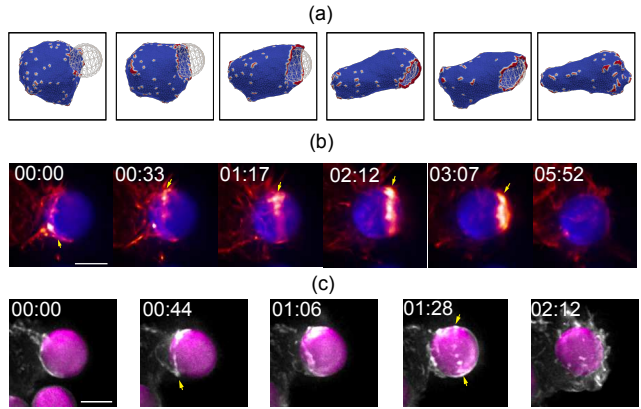}
\caption{Comparison of the actin organization at the rim of the phagocytic cup, observed in simulations (a) and experiments (b,c). In both the experiments and the simulations we find that the actin ring is highly fragmented, in the form of dispersed "teeth" and "arcs" along the leading edge (yellow arrows in (b,c)). The actin aggregate becomes more cohesive in the final stages of the engulfment, before it disperses after the engulfment is complete. (a) Snapshots from a simulation, using $E_{ad}=1.50 ~k_BT$, $\rho=6.4~\%$ and $F=1.0 ~k_BT/l_{min}$. (b,c) Time-lapse sequences of maximum intensity projection images illustrating engulfment of Immunoglobulin-G-coated polystyrene beads ($7~\mu m$ diameter) by RAW264.7 macrophage-like cell line (b) and by murine bone-marrow derived macrophages (c), imaged by lattice-light sheet microscopy. In b and c, cells were transfected with mEmerald-Lifeact to label F-actin. Scale bar $5~\mu m$, time is indicated in min:sec.}
\label{fig:experiment}
\end{figure}

\begin{figure}[h]
\centering
\includegraphics[scale=0.65]{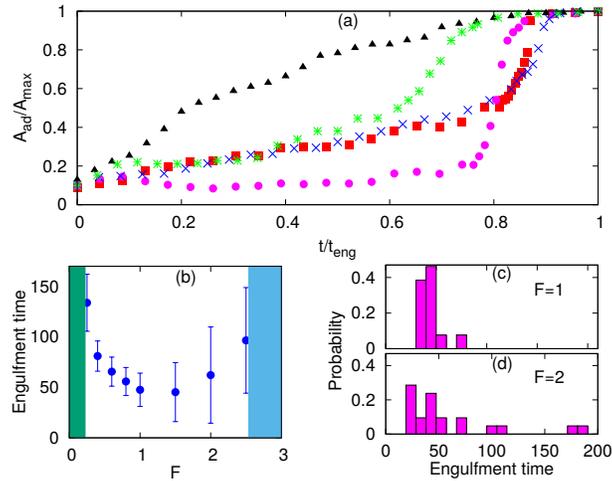}
\caption{Engulfment dynamics for spherical particles. (a) Different growth behaviour of adhered fraction with time for protein-free, passive and active cases. We scale the area axis by maximum adhered area ($A_{max}$) and time axis by the engulfment time ($t_{eng}$). Blue cross symbols are for protein free case with $E_{ad}=1.30 ~k_BT$; green star symbols for passive proteins with $E_{ad}=1.0 ~k_BT$ and $\rho=4.0 ~\%$; red boxes for $E_{ad}=1.0~ k_BT$, $\rho=1.6 ~\%$ and $F=0.40~ k_BT/l_{min}$; magenta circles for $E_{ad}=1.0~k_BT$, $\rho=1.6 ~\%$ and $F=2.0 ~k_BT/l_{min}$; black triangles for $E_{ad}=1.50 ~k_BT$, $\rho=6.4 ~\%$ and $F=1.0 ~k_BT/l_{min}$. (b) Engulfment time as function of $F$, showing the non-monotonic behavior: below and above a critical force (green and blue shaded areas respectively) there is partial engulfment. (c-d) Distributions of the engulfment time for $F=1,2 ~k_BT/l_{min}$, respectively.}
\label{fig:eng-dynamics}
\end{figure}

However, for the active system at high protein density, where the rim clusters form at an early stage of the engulfment process (Fig. \ref{fig:active}e), we find a much more uniform dynamics with steady growth of the adhered fraction (Fig. \ref{fig:eng-dynamics}a). This agrees qualitatively with the observations in \cite{barger2019membrane,Nils2021}, which indeed correspond to this regime (Figs. \ref{fig:active}e,\ref{fig:experiment}).

In Fig. \ref{fig:eng-dynamics}a the engulfment dynamics are shown as function of normalized time, so the actual engulfment duration is not presented. The dependence of the mean engulfment time on the active force is found to be non-monotonic (Fig. \ref{fig:eng-dynamics}b). Below a critical force, in the regime of low $E_{ad}$ and low $\rho$, where the passive system does not fully engulf (Fig. \ref{fig:active}a), there is no complete engulfment (green shaded area). Above this regime, increasing force induces complete engulfment with faster and more robust engulfment as $F$ increases. The engulfment time distribution in this regime is relatively narrow (Fig. \ref{fig:eng-dynamics}c). 

However, increasing $F$ beyond some value, which is set by the relation of the adhesion energy and the active work, leads to increasing both the mean value and the variability of the engulfment time. This occurs due to active forces stretching the membrane away from the adhered particle (top insets in Fig. \ref{fig:active}c), which gives rise to long "waiting times" during which the engulfment does not progress. Following this "waiting time", once the protein aggregate forms at the membrane rim (Fig. \ref{fig:active}c), the engulfment proceeds rapidly as seen in Fig. \ref{fig:eng-dynamics}a (magenta circles). This variability is manifested in the very broad distribution of engulfment times (Fig. \ref{fig:eng-dynamics}d).

At even larger forces, the stretching of the membrane sideways by the active forces prevents engulfment (blue shaded area in Fig. \ref{fig:eng-dynamics}b, and SI sec S8, Fig. S12(a), Movie-S14). Large membrane tension is indeed known to hinder and prevent engulfment in cells \cite{masters2013plasma}. In this regime of large $F$, if we further increase $\rho$, we reach the regime where a free vesicle will form a flat pancake-like shape, where all the proteins are aggregated at the rim \cite{miha2019}. In this regime the vesicle can engulf the particle if this process is faster than the timescale of the formation of the pancake-like shape (SI sec. S8, Fig. S12(b); Movie-S15). For very large forces, we find again a non-engulfed regime (similar to the blue shaded area in Fig. \ref{fig:eng-dynamics}b, SI Fig. S12(c), Movie-S16).

\subsection{Engulfment of non-spherical particles}
Immune cells often engulf highly non-spherical objects, such as bacteria . Furthermore, studies with different particle shapes indicated a strong shape-dependence of the engulfment time and success rate \cite{Champion2006,Jeroen2009,Richards2016}. We therefore study the engulfment of non-spherical particles by our vesicles that contain passive or active curved proteins. 

We first consider spheroids. We keep the surface area constant, i.e. same as for the sphere of radius $10 ~l_{min}$. We use prolate shapes ($R_x = R_y < R_z$) and oblate shapes ($R_x = R_y > R_z$), such that the aspect ratio $R_x/R_z < 1$ for prolate, and $R_x/R_z > 1$ for oblate. 

\begin{figure}[h!]
\centering
\includegraphics[scale=0.7]{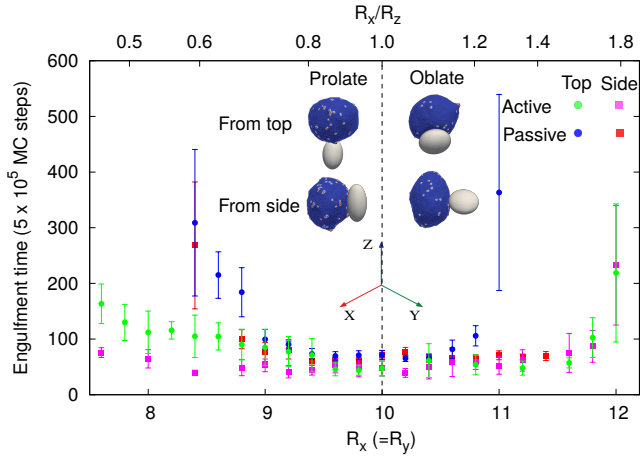}
\caption{Mean engulfment time for non-spherical particles. The vertical dashed line is separating oblate (right) and prolate (left) shapes. Here, for the passive case, we use: $E_{ad}=1.0~k_BT$, $\rho=4.8~\%$, and for the active case: $E_{ad}=1.0~k_BT$, $\rho=1.6~\%$, and $F=1.0~k_B T/l_{min}$. The surface area of all the shapes is constant, and equal to that of a sphere of radius $10~l_{min}$. The insets demonstrate the initial conditions of either starting from the top (poles) of the shapes (along their axis of rotational symmetry), or from the side (equator).}
\label{fig:shperoid}
\end{figure}

In Fig. \ref{fig:shperoid}, we show the mean engulfment time for oblate and prolate spheroids, for both the cases when the vesicle is initially in contact from the top (the poles of the shape) or from the side (along the equator), as shown in the insets. For both passive and active curved proteins, we find that the engulfment times increase as the shape deviates from a sphere. This is in general agreement with previous experimental observations \cite{Champion2006}, and can be intuitively attributed to the higher curvature of the non-spherical shapes, which increase the bending energy cost of their engulfment.

For the particles with small aspect ratio ($R_x/R_z$ close to unity), the engulfment time for all the cases are comparable. However, for the particles with large aspect ratios, there are significant differences in the engulfment times, between the passive and active systems. For passive curved proteins (blue circles and red boxes), we find that the engulfment times increase sharply with the deviation from spherical shape. Furthermore, we find that initialising from the side (red boxes) results in relatively shorter engulfment times, especially for the oblate shapes. For an oblate particle with large aspect ratio (for $R_x=R_y>11~l_{min}$), the particle is engulfed from side, while it is not engulfed from the top (Fig. \ref{fig:shperoid}). For prolate shapes with large aspect ratios, however, the engulfment times seem comparable for both initial conditions. 

By comparison, for active curved proteins the engulfment time is lower than for the passive case, even with a lower protein density, and the active curved proteins enable complete engulfment over a larger range of aspect ratios. We also note that for the active curved proteins it is the prolate shape that exhibits a dependence of the engulfment time on the initial conditions, where starting from the side (magenta boxes) leads to faster engulfment. For the oblate shapes, however, the engulfment time is largely independent of the initial conditions, opposite to the behavior of the passive curved proteins. 

For a non-spherical particle with high aspect ratio, due to the highly curved poles (in the prolate) or equator (oblate) regions, the engulfment process is slower and the variability in engulfment times increases. The slower engulfment allows the vesicle to reorient on the surface of the particle, driven by both passive energy minimization and active work. We show the details of this reorientation process in the SI sec. S9 (Movie-S17-S22).

When the vesicle is initially at the top of a prolate shape with high aspect ratio, the vesicle reorients itself always to the side and engulfs the particle. For passive curved proteins, this reorientation process is faster and thereby the engulfment time when started from top or from side are comparable (blue circles and red boxes in Fig. \ref{fig:shperoid} for $R_x/R_z < 1$). For active curved proteins, this reorientation process is slower because of large shape fluctuations of the vesicle, and thus the engulfment time when started from the side is smaller than from the top (magenta boxes in Fig. \ref{fig:shperoid} for $R_x/R_z < 1$).

For the oblate shapes we find a similar reorientation process when the vesicle starts the adhesion from the highly curved sides. In this geometry, the passive vesicle can engulf from side without any reorientation, which is very fast, or it reorients to the top of the oblate shape and ends up with no engulfment. Therefore, when we calculate the engulfment time we only average over those realizations where the particle was engulfed without any reorientation (red boxes for $R_x>11$). For the active case, the vesicle reorients in every realization and engulfs from the top. This reorientation process is very fast and the engulfment time is similar for starting either from the side or from the top (green circles and magenta boxes in Fig. \ref{fig:shperoid} for $R_x/R_z > 1$).

We further explore more extreme non-spherical shapes, such as elongated sphero-cylinders and dumb-bell shapes in the SI (Sec. S10, Movie-S23-S26). The elongated particles resemble the shapes of different bacteria or fungi such as \textit{E. coli}, \textit{Bacillus subtilis} or budding yeast \cite{moller2012race,Richards2016,Clarke2010}.

\section{Discussion}
In this study we demonstrated that curved (convex) membrane proteins, both passive and active, can strongly affect the engulfment process of rigid particles by an adhesive vesicle. The engulfment process that we calculate, especially using active curved proteins, exhibits many features that are observed during phagocytosis of rigid particles and pathogens by living cells. We therefore propose that our model provides a mechanism for the self-organization of the membrane and the actin cytoskeleton in the phagocytic cup.

We find that the curved proteins spontaneously aggregate around the highly curved rim of the phagocytic cup, and enable the engulfment to proceed efficiently by reducing the bending energy costs. The active forces in our model, which represent the protrusive forces due to the recruited actin polymerization, further facilitate the engulfment, due to several effects that we have identified: the protrusive force at the membrane leading edge promotes the aggregation of the curved proteins (reducing the bending energy cost), as well as directly pushing the membrane over the engulfed particle. In addition, large forces can stretch the membrane to wrap around sharp corners, which enables engulfment of non-spherical particles. Compared to the passive curved proteins, the curved active proteins drive more efficient engulfment at lower adhesion energy, lower curved protein density, and of highly non-spherical particles. These results of our model highlight the central role that the recruitment of actin plays during phagocytosis, which is usually abolished by inhibition of actin polymerization (see for example \cite{prashar2013filamentous}).

Our model predicts that the actin aggregation at the rim of the phagocytic cup is typically fragmented and does not form a complete circular ring. Nevertheless, it is still highly effective in driving the engulfment. Furthermore, the actin aggregate forms a more cohesive ring in the final stages of the engulfment, before it spontaneously disperses as the engulfment is complete.  These features of our model are verified and observed in recent high-resolution imaging of the actin during phagocytosis \cite{Nils2021} (Fig. \ref{fig:experiment}). 

The model we presented currently lacks the contractile forces that arise at the phagocytic cup due to myosin-II activity \cite{Nils2021}. The contractile forces play an important role during the engulfment of soft objects, such as cells \cite{tsai2008inhibition,10.1242/jcs.00235,barger2020squeezing}, but they are dispensable for the phagocytosis of more rigid particles \cite{prashar2013filamentous,Nils2021}. Contribution of myosin-II-mediated contractility may be more or less pronounced depending on the particle size and stiffness and, in addition, could influence mechanosensitive signaling within phagocytes rather than directly drive phagocytic cup extension and closure. Our results should therefore be important even without this feature, which we plan to add to our model in the future, especially when modelling the engulfment of flexible particles.

To conclude, we demonstrate that curved membrane proteins can play a key role during phagocytosis. Convex-shaped proteins or a curved complex of proteins, that furthermore recruit actin polymerization, spontaneously aggregate at the leading edge of the membrane, and drive robust engulfment. This is similar to the role of these proteins that was recently found at the leading edge of the lamellipodia, in experiments\cite{begemann2019mechanochemical,pipathsouk2021wave} and theory \cite{Sadhu2021}. We also show that concave-shaped proteins can further enhance the engulfment process by reducing the bending energy cost of the membrane that adheres to the particle's surface (SI sec. S11, Fig. S18, Movie-S27). Our results should motivate future experimental studies aimed at the identification and characterization of the different curved membrane components that participate in phagocytosis \cite{flannagan2012cell,bohdanowicz2013role,joern2014}.

\section{Materials and Methods}
\subsection{Theoretical modelling}
The membrane shape and protein positions are simulated using a Monte-Carlo (MC) procedure \cite{Sadhu2021,miha2019}. The vesicle energy, which is calculated per each MC move, has the following contributions: (1) The bending energy, which depends on the bending rigidity $\kappa$, local mean curvature $H$ and the spontaneous curvature $C_0$ (which we take to be nonzero for the nodes occupied by the membrane proteins), (2) The attractive protein-protein interaction energy between nearest-neighbors, of strength $w$, (3) The magnitude of the force of the active force $F$, pushing towards the outward normal vector of the vertex that contains a protein, and finally, (4) The adhesion energy of magnitude $E_{ad}$ per node that is within close proximity to the external substrate \cite{samo2015,miha2019,Sadhu2021}. For more details, please see SI sec. S1.

Unless specified, we use a vesicle of total number of vertices, $N=3127$ (radius $\sim 20~ l_{min}$), where $l_{min}$ is the unit of length in our model, and defines the minimum bond length. The bending rigidity $\kappa =20 ~k_B T$, the protein-protein attraction strength $w=1 ~k_BT$, and $\rho=N_c/N$ is the protein density, with $N_c$ vertices occupied by curved (convex) membrane proteins having spontaneous curvature: $C_0=1.0~ l_{min}^{-1}$. Note that we do not conserve the vesicle volume, although this could be maintained using an osmotic pressure term \cite{miha2019}. The membrane area is very well conserved.

\subsection{Imaging of Macrophage Phagocytosis}
Culturing of RAW264.7 macrophages (ATCC) and generation of murine bone-marrow derived macrophages is described in \cite{barger2019membrane}. Macrophages were transfected with mEmerald-Lifeact (Addgene, $\# 54148$) to label F-actin using the Neon electroporation system, according to manufacturer’s instructions, and allowed to recover for $24~hr$.  Flash Red polystyrene beads (Bangs Laboratories Inc., $7~ \mu m$ diameter) were washed three times in sterile PBS and opsonized overnight at $4^{\circ}$C in $3~mg/mL$ mouse IgG (Invitrogen). To remove excess antibody, beads were washed three times with PBS and resuspended in sterile PBS. Beads were applied to macrophages, plated on $5~mm$ round coverslips, in a $37^{\circ}$C-heated, water-coupled bath in imaging medium (FluoroBrite (Thermo Scientific), $0-5~\%$ FBS, Pen/Strep) prior to imaging acquisition. Imaging was performed using a lattice light-sheet microscope operated and maintained by the Advanced Imaging Center at the Howard Hughes Medical Institute Janelia Research Campus (Ashburn, VA). $488,~ 560$, or $642~ nm$ diode lasers (MPB Communications) were operated between $40$ and $60~mW$ initial power, with $20-50~\%$ acousto-optic tunable filter transmittance. The microscope was equipped with a Special Optics $0.65~NA/3.75~mm$ water dipping lens, excitation objective, and a Nikon CFI Apo LWD $25 \times 1.1~NA$ water dipping collection objective, which used a $500~mm$ focal length tube lens. Images were acquired with a Hamamatsu Orca Flash $4.0$ V2 sCMOS cameras in custom-written LabView Software. Post-image deskewing and deconvolution was performed using HHMI Janelia custom software and $10$ iterations of the Richardson-Lucy algorithm.



\section{Acknowledgment}
We acknowledge useful comments by Orion Weiner. N.S.G. is the incumbent of the Lee and William Abramowitz Professorial Chair of Biophysics, and acknowledges support by the Ben May Center for Theory and Computation. A.I. and S.P. were  supported by the  Slovenian Research Agency (ARRS) through the Grant No. J3-3066 and  Programme No. P2-0232. LLSM imaging was performed at the Advanced Imaging Center (AIC)-Howard Hughes Medical Institute (HHMI) Janelia Research Campus. The AIC is jointly funded by the Gordon and Betty Moore Foundation and the Howard Hughes Medical Institute.

\section{Supplementary movies}
All the supplementary movies are available online in the link: https://app.box.com/s/ske4vjlp6ci3vu7mz5izfp2iebxrmgm4

\pagebreak

\setcounter{equation}{0}
\renewcommand{\thefigure}{S-\arabic{figure}}
\renewcommand{\thesection}{S-\arabic{figure}}
\renewcommand{\theequation}{S-\arabic{figure}}
\setcounter{section}{0}
\setcounter{figure}{0}

\section{Supplementary informations}

\subsection{Model and simulation details}\label{sec:model}
\begin{figure}[ht]
\centering
\includegraphics[scale=0.7]{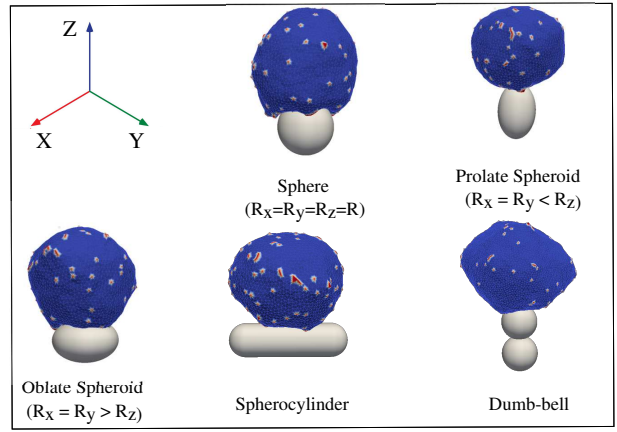}
\caption{Schematic representation of our model. The vesicle is kept in contact with a particle, having attractive interaction between them. We consider spherical as well as non-spherical particles, such as spheroid, spherocylinder, dumb-bell etc.}
\label{fig:model}
\end{figure}

We consider a three-dimensional vesicle, which is described by a closed surface having $N$ vertices, each of them connected to its neighbours with bonds, to form a dynamically triangulated, self-avoiding network, with the topology of a sphere \cite{Sadhu2021,miha2019}. The vesicle is kept in contact with a particle, with which the vesicle has an attractive interaction, as shown in Fig. \ref{fig:model}. The vesicle energy has four following contributions: The bending energy is given by,

\begin{equation}
    W_b=\frac{\kappa}{2} \int_A (C_1 + C_2 - C_0)^2 dA,
\end{equation}

where $\kappa$ is the bending rigidity, $C_1$, $C_2$ are the principal curvatures and $C_0$ is the spontaneous curvature. The nearest neighbour protein-protein direct interaction energy is given by, 

\begin{equation}
    W_d = -w\sum_{i<j} {\cal H} (r_0 - r_{ij}),
\end{equation}

where, ${\cal H}$ is the Heaviside step function, $r_{ij}=|\overrightarrow{r}_j - \overrightarrow{r}_i|$ is the distance between proteins, $\overrightarrow{r}_i, \overrightarrow{r}_j$ are the position vectors for $i,j-th$ proteins, and $r_0$ is the range of attraction, $w$ is the strength of attraction. The range of attraction is such that only the proteins that are in the neighbouring vertices can attract each other. The active energy is given by, 

\begin{equation}
    W_F = -F \sum_i \hat{n_i}.\overrightarrow{r}_i,
\end{equation}

where, $F$ is the magnitude of the force representing protrusive force due to actin polymerization, $\hat {n_i}$ is the outward normal vector of the vertex that contains a protein and $\overrightarrow{r}_i$ is the position vector of the protein. Finally, the adhesion energy is given by, 

\begin{equation}
    W_A = -\sum_{i'} E_{ad},
\end{equation}

where $E_{ad}$ is the adhesion strength, and the sum runs over all the vertices that are adhered to the phagocytic particle \cite{samo2015,miha2019,Sadhu2021}. By `adhered vertices', we mean all such vertices, whose perpendicular distance from the surface of the particle are less than $l_{min}$. Thus, the total energy of the vesicle-particle system is given by,

\begin{equation}
    W = W_b + W_d + W_F + W_A
\end{equation}

We update the vesicle with mainly two moves, (1) vertex movement and (2) Bond flip. In a vertex movement, a vertex is randomly chosen and attempt to move by a random length and direction, with the maximum possible distance restricted by $0.15 ~l_{min}$. In a bond flip move, a single bond is chosen,  which is a common side of two neighbouring triangles, and this bond is cut and reestablished between the other two unconnected vertices \cite{samo2015,miha2019,Sadhu2021}. The maximum bond length is restricted to $l_{max}=1.7 ~l_{min}$. The particle is assumed to be static and does not have any independent dynamics. We update the system using Metropolis algorithm, where any movement that increases the energy of the system by an amount $\Delta W$ occurs with rate $exp(- \Delta W/k_BT)$, otherwise it occurs with rate unity. Note that our model only captures the wrapping and engulfment events, while the other upcoming processes, such as the membrane fission, and the movement of the engulfed particle inside the vesicle are not captured in the present model.

\subsection{Results for protein-free vesicle}
\begin{figure}[ht]
\centering
\includegraphics[scale=0.9]{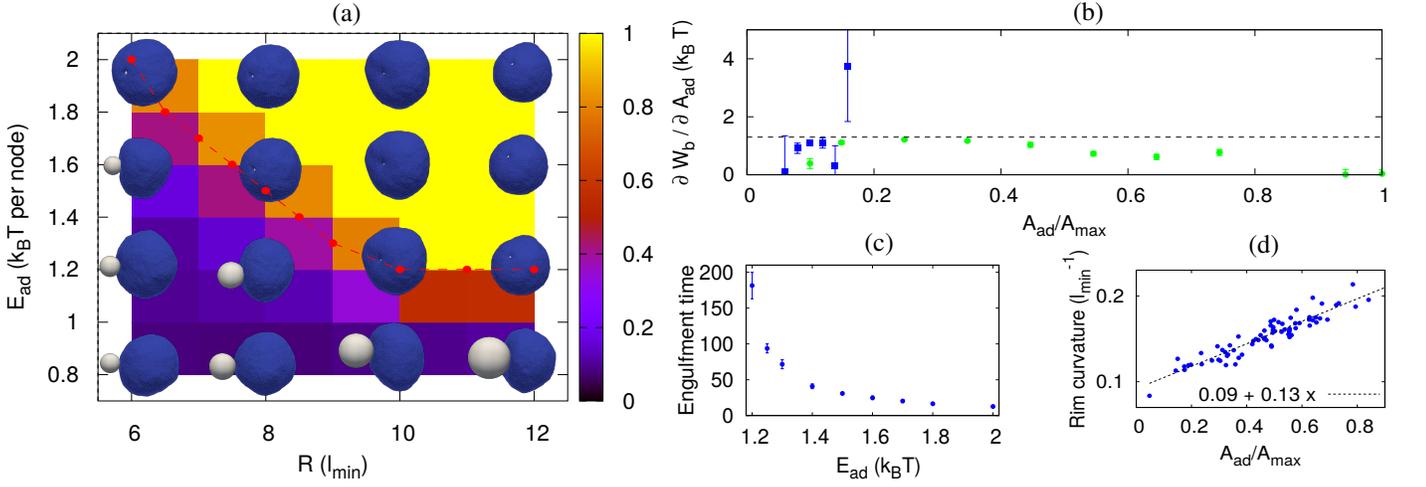}
\caption{Engulfment of spherical particles by a protein-free vesicle. (a) Phase-diagram in the $R-E_{ad}$ plane. The background color is showing the adhered fraction. (b) Comparison of adhesion energy gain and bending energy cost (both in units of $k_BT$) for partial and complete engulfment. Blue boxes are for $R=8~ l_{min}$ and green circles are for  $R=10~l_{min}$. We use $E_{ad}=1.30~k_BT$ here. (c) Engulfment time with $E_{ad}$ showing divergence near the critical $E_{ad}~(\sim 1.20~k_BT)$ below which there is no engulfment. Here, we use $R=10~l_{min}$. The unit of time is $5 \times 10^5$ M C steps. (d) Mean curvature at the rim (the leading edge of the vesicle) with adhered fraction $A_{ad}/A_{max}$ for a complete engulfment. The parameters are same as in Fig. \ref{fig:protein-free}b, green circles.}
\label{fig:protein-free}
\end{figure}

Here, we discuss the engulfment of spherical particles by a protein-free vesicle. In Fig. \ref{fig:protein-free}a, we show the phase diagram in the $R-E_{ad}$ plane. The background color is showing the adhered fraction $A_{ad}/A_{max}$, where $A_{ad}$ is the adhered area and $A_{max}$ is the maximum adhered area. The value of $A_{max}$ is in general larger than the surface area of the particle, since we have a finite width of adhesive interaction ($l_{min}$), within which the vesicle can fluctuate. This value may also be different for different parameter regime, but the difference is almost negligible. In the calculation of adhered area fraction (background color of Fig. \ref{fig:protein-free}a) we implement the following method: For a complete engulfment, we assume $A_{max}$ to be the maximum adhered area for a given $E_{ad}$ with different values of $\rho$, and divide all the adhered area by that value of $A_{max}$ to get the adhered area fraction. For a partial engulfment, we choose the $A_{max}$ corresponding to a higher $E_{ad}$ closest to the original value of $E_{ad}$, for which we have a complete engulfment. We use similar methods for calculating the adhered area fraction for the passive and active cases as well.

We note that smaller particles are difficult to engulf, as the bending energy cost becomes higher than the adhesion energy gain (Movie-S2). So, in order to engulf a small particle, $E_{ad}$ should be increased to large values. On the other hand, larger particle is engulfed at smaller $E_{ad}$ (Movie-S1). Since the size of the vesicle is finite, a much larger particle will not be engulfed by the vesicle, rather the vesicle will spread over it. So, there is an optimal size of the particle, that requires the smaller $E_{ad}$ to engulf. We consider this optimal radius to be $R=10~l_{min}$ and use this value in the main paper.

For a given values of $E_{ad}$ we choose two particles with different sizes $R$, such that for one value, the engulfment is partial, while for the other value, the engulfment is complete. We then compare the adhesion energy gain and bending energy cost for both the cases in Fig. \ref{fig:protein-free}(b). Similar to the previous cases (passive and active vesicle), here also, we can explain the engulfment process by the competition between adhesion and bending energy costs.

We also measure the engulfment time for a given $R(=10 ~l_{min})$ and for various $E_{ad}$. The engulfment time shows a divergence as we reach the critical $E_{ad}$ (below which engulfment is incomplete) (Fig. \ref{fig:protein-free}c). We also measure the curvature of the vesicle at the leading edge (rim) with the adhered fraction $A_{ad}/A_{max}$ as the engulfment proceeds. We note that the curvature at the leading edge increases linearly with $A_{ad}/A_{max}$ (Fig. \ref{fig:protein-free}d). We use this linear behaviour of rim curvature in our analytical calculations.

\subsection{Analytical model for protein-free vesicle}
\begin{figure}[ht]
\centering
\includegraphics[scale=0.7]{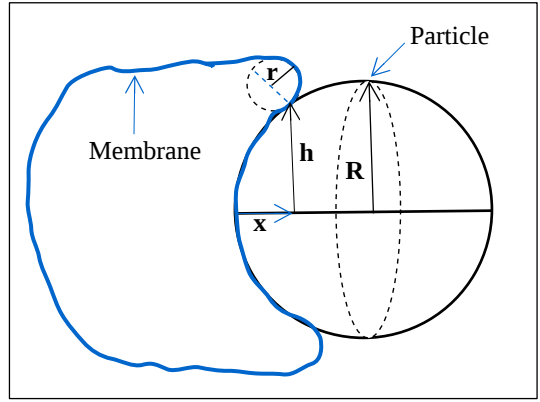}
\caption{Schematic diagram of our analytical model. A spherical particle with radius $R$ is being engulfed by a cell membrane of size $R_0$. We show the projection of the system on a plane. }
\label{fig:analytic}
\end{figure}

Here, we present our simplified analytical (parametrical) model for protein free case. We consider a three-dimensional vesicle formed by a closed two-dimensional membrane, adhering to a spherical particle. We assume the system to be symmetric, such that we can only show a two-dimensional projection of the system on a plane (Fig. \ref{fig:analytic}). The vesicle area is $A=4 \pi R_0^2$, and the particle radius is $R$, such that $R<R_0$. As the vesicle engulfs the particle, the adhered fraction increases. The adhered area can be written as, $A_{ad}=2 \pi x R$, where $x$ is the thickness of the spherical cap of the particle that is adhered, as shown in Fig. \ref{fig:analytic}. The leading edge of the membrane (rim) is highly curved, and the shape is assumed to be similar to a half-torus with one principal curvature $1/r$ and the other $1/h$. We assume $r \ll h$ and hence the other principal curvature is neglected. The area of this region can be written as $A_{rim}=2 \pi^2 r h$, where $h=\sqrt{2Rx-x^2}$.

The bending energy of the adhered section of the membrane is, 
\begin{equation}
    W_{b1}=\frac{\kappa}{2} A_{ad} (2/R)^2
\end{equation}

and, the bending energy of the rim section is,

\begin{equation}
    W_{b2}=\frac{\kappa}{2} A_{rim} (1/r)^2
\end{equation}
where, the mean curvature at the rim has the form, $1/2r = 1/R_0 + \alpha x/R$ (from simulation data in Fig. \ref{fig:protein-free}d, where $\alpha$ is a fitting parameter), where $\alpha$ a constant and $A_{ad}/A \sim x/2R$, such that at $x=0$, mean curvature becomes $1/R_0$, which is the mean curvature of the unadhered spherical vesicle.

We assume that the rest part of the unadhered vesicle have the same mean curvature as before, $i.e.,$ $1/R_0$, thus, the bending energy of this section will be,

\begin{equation}
    W_{b3}=\frac{\kappa}{2} (A-A_{rim} - A_{ad}) (2/R_0)^2
\end{equation}

The total bending is thus given as,

\begin{equation}
    W_b = W_{b1} + W_{b2} + W_{b3}
\end{equation}

We can now compare the adhesion energy gain and bending energy cost, as the engulfment proceeds, and can calculate the transition line.

\subsubsection{Graphical solution for the transition line in the $R-E_{ad}$ plane}
\begin{figure}[ht]
\centering
\includegraphics[scale=0.5]{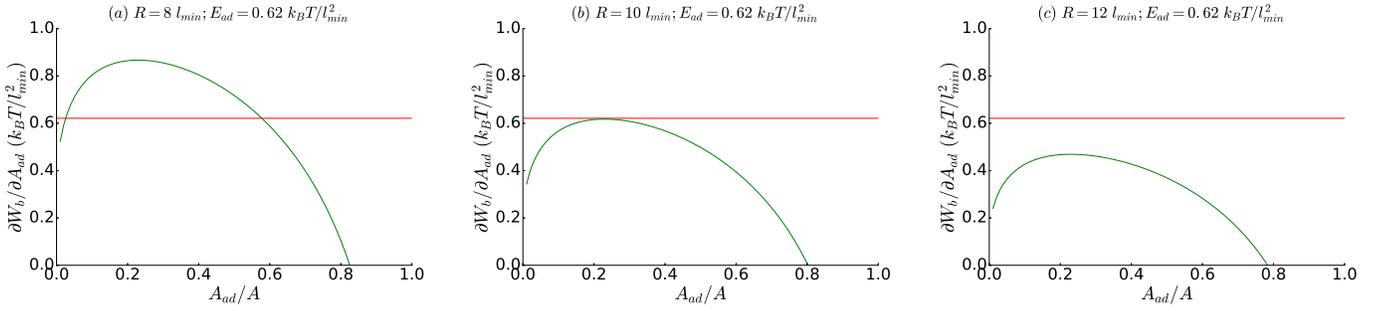}
\caption{Comparison of adhesion energy gain (red line) and bending energy cost (green line) for few values of $R$ ($=8,10,12~l_{min}$) and given $E_{ad}=0.62~k_BT/l_{min}^2$. Other parameters are, $R_0=20~l_{min}$, $\alpha=0.065$, $\kappa=20~k_BT$.}
\label{fig:analytical-no-protein}
\end{figure}

For a given value of $E_{ad}$, we plot the adhesion energy gain ($E_{ad}$) and the bending energy cost $\partial W_b/\partial A_{ad}$ with adhered area fraction, for few values of $R$ (Fig. \ref{fig:analytical-no-protein}). For very small $R$, the adhesion strength is not enough to overcome the bending energy cost and the particle will not be engulfed (Fig. \ref{fig:analytical-no-protein}a). For a critical value of $R$, the adhesion strength will just be sufficient to overcome the bending energy cost, and the particle will be engulfed, as shown in Fig. \ref{fig:analytical-no-protein}b. If we further increase $R$, the adhesion energy gain will be always higher than the bending energy cost, and the particle will easily be engulfed  (Fig. \ref{fig:analytical-no-protein}c). 

\begin{figure}[ht]
\centering
\includegraphics[scale=0.7]{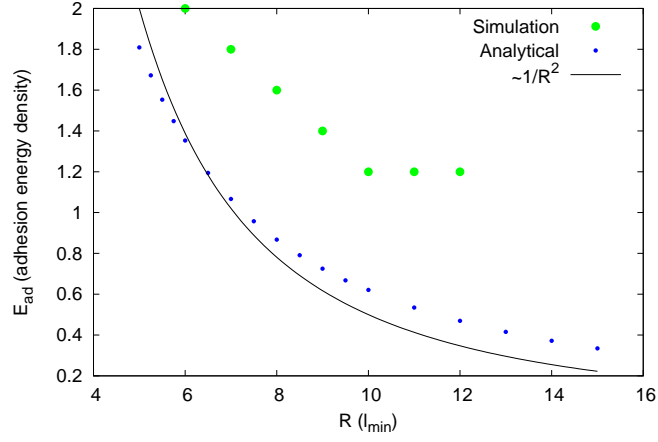}
\caption{Transition line for engulfment in the $R-E_{ad}$ plane and comparison with simulations. In our analytical calculations, we define $E_{ad}$ as the adhesion energy per unit area, while in simulation, we define it as the adhesion energy per adhered node. So, while comparing, we scale $E_{ad}$ properly, so that the definition remains consistent with each other. The green circles are simulation result. The analytical prediction for the critical value of $R$ with $E_{ad}$ (blue circles), which is close to the classical prediction of $E_{ad} \sim 1/R^2$ \cite{Lipowsky1998}, shown by black solid line. For simulation data, the parameters are same as in Fig. \ref{fig:protein-free}a, and for analytical results, parameters are same as Fig. \ref{fig:analytical-no-protein}.}
\label{fig:E-R-compare}
\end{figure}

We calculate the transition line by finding the line corresponding to $E_{ad}$ which is just touching the peak of bending energy cost $\partial W_b/\partial A_{ad}$. We show this transition line in Fig. \ref{fig:E-R-compare}, and also compare with the simulation line and the classical result from Ref. \cite{Lipowsky1998}. For the analytical model, we define $E_{ad}$ as the adhesion energy per unit area, while in our simulation, we define it as the adhesion energy per vertex. While comparing the simulation and analytical results, we properly scale $E_{ad}$ such that they become consistent with each other.

\subsection{Extension of the analytical model for passive curved (convex) proteins case}
We here extend our analytical model for the passive curved protein case, with spontaneous curvature $c_0=1.0~l_{min}^{-1}$. Let $\rho_0$ be the average protein density on the membrane, $\rho_r$ be the density at the rim and $\rho_s$ be the density on the rest of the unadhered part of the vesicle. We assume there are no proteins on the membrane that is adhered to the spherical substrate. Thus,

\begin{equation}
    A \rho_0 = A_{rim} \rho_r + (A - A_{rim} - A_{ad}) \rho_s
\end{equation}

The bending energy of the adhered section is given by,

\begin{equation}
    W_{b1}=\frac{\kappa}{2} A_{ad} (2/R)^2
\end{equation}

The bending energy of the rim is,

\begin{equation}
    W_{b2}=\frac{\kappa}{2} A_{rim} (1/r - c_0 \rho_r)^2
\end{equation}

where, the mean curvature at the rim, $1/2r = 1/R_0 + \alpha x/R$ and $\rho_r = \rho_0 + \beta x/R$ is the density at the rim (validated from simulation, Fig. \ref{fig:rim-passive}), where $\alpha=0.093$ and $\beta=0.085$ (determined from Fig. \ref{fig:rim-passive}). Assuming that the rest part of the unadhered vesicle have the same curvature of $1/R_0$, the bending energy of this part of the vesicle is given by,

\begin{equation}
    W_{b3}=\frac{\kappa}{2} (A-A_{rim} - A_{ad}) (2/R_0 - c_0 \rho_s)^2
\end{equation}

Thus, the total bending energy cost is given by,

\begin{equation}
    W_b = W_{b1} + W_{b2} + W_{b3}
\end{equation}

\subsubsection{Mean curvature and protein density at the rim (from simulation)}
\begin{figure}[ht]
\centering
\includegraphics[scale=1.4]{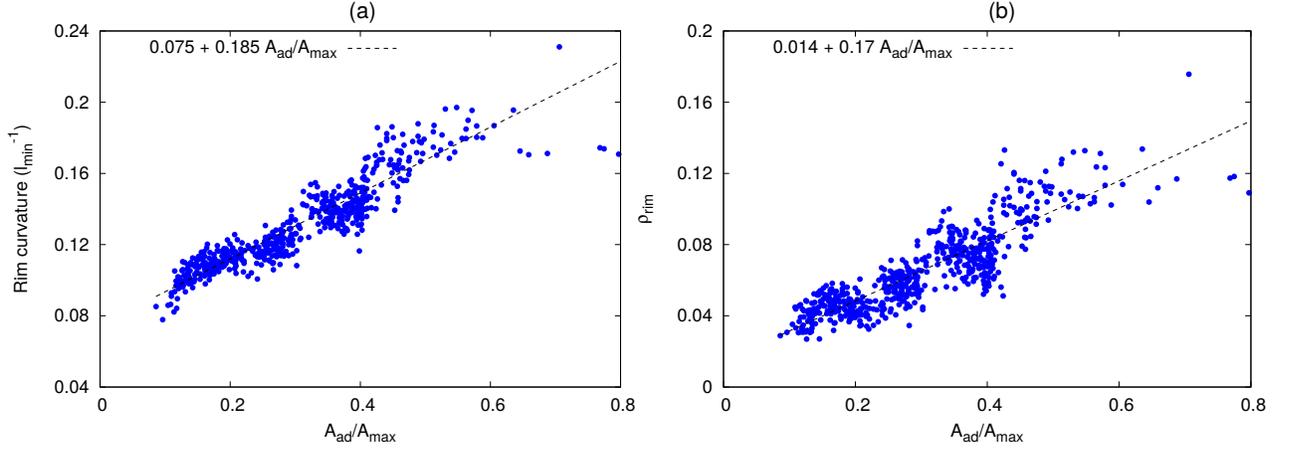}
\caption{Mean curvature and protein density at the highly curved rim for passive case. (a) The mean curvature at the rim with adhered fraction of the particle $A_{ad}/A_{max}$. (b) The protein density at the rim with $A_{ad}/A_{max}$. The parameter used here are, $R=10~ l_{min}$, $E_{ad}=1.04~ k_BT$, and $\rho=3.2~\%$. }
\label{fig:rim-passive}
\end{figure}

For the passive case with small protein density, we determine the mean curvature of the vesicle at the rim and plot it with adhered fraction of the particle, $A_{ad}/A_{max}$ in Fig. \ref{fig:rim-passive}a. Similar to the protein-free case, here we note that the rim curvature seems to increase linearly, with a little higher slop than the protein-free case. We also determine the protein density at the rim, and plot it with adhered fraction of the particle, $A_{ad}/A_{max}$ (Fig. \ref{fig:rim-passive}b). The rim density also seems to vary linearly with the adhered fraction. 

\subsubsection{Graphical solution for the engulfment transition line in the $\rho-E_{ad}$ plane}
\begin{figure}[ht]
\centering
\includegraphics[scale=0.5]{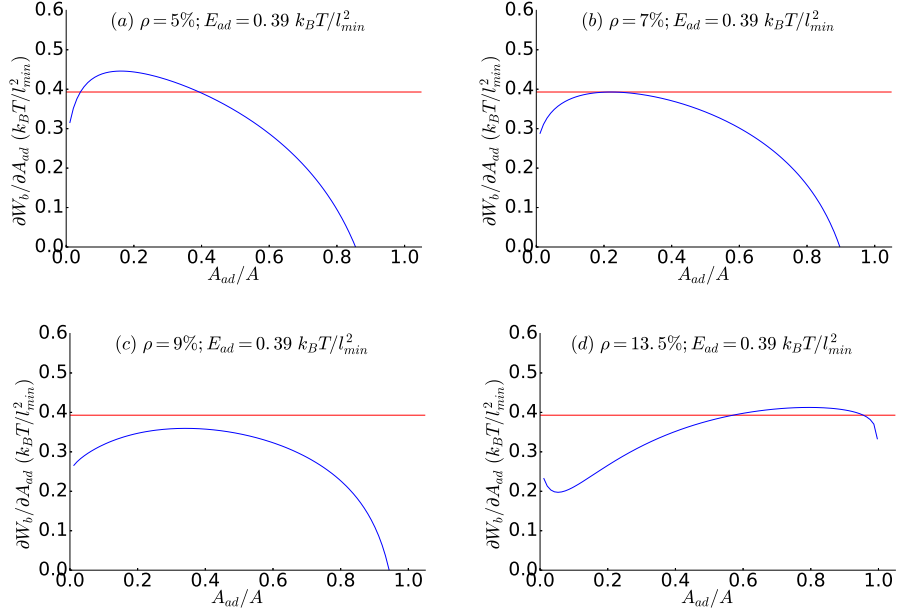}
\caption{Comparison of adhesion energy gain (red line) and bending energy cost (green line) for few values of $\rho$ ($=5~\%, 7~\%, 9~\%, 13.5~\%$) and given $E_{ad}=0.39~k_BT/l_{min}^2$. Other parameters are, $R=10~l_{min}$, $R_0=20~l_{min}$, $\alpha=0.093$, $\beta=0.085$, $\kappa=20~k_BT$.}
\label{fig-analytical-passive}
\end{figure}

\begin{figure}[ht!]
\centering
\includegraphics[scale=0.7]{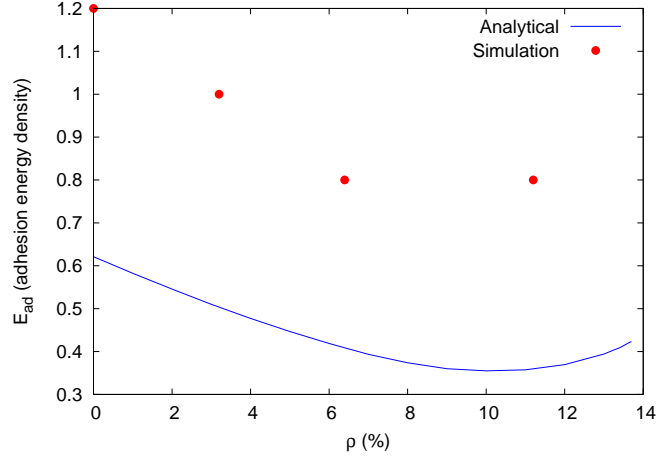}
\caption{Transition line for engulfment in the $\rho-E_{ad}$ plane and comparison with simulations. The red circles are simulation result, and the blue line is the analytical prediction. For the simulation results, the parameters are the same as in Fig. 1(a) of the main text. For analytical results, parameters are the same as in Fig. \ref{fig-analytical-passive}.}
\label{fig:passive-compare}
\end{figure}

Similar to the protein-free case, here also, we calculate the transition line in the $\rho-E_{ad}$ plane by using graphical method. We compare the adhesion energy gain and bending energy cost with adhered area fraction, for few values of protein density $\rho$ and a given $R=10~l_{min}$ and given $E_{ad}=0.39~k_BT/l_{min}^2$ in Fig. \ref{fig-analytical-passive}. For a small $\rho$, the vesicle will remain unadhered (Fig. \ref{fig-analytical-passive}a). As we increase $\rho$, the adhesion energy gain will just become sufficient to overcome the bending energy cost, which is the transition value (Fig. \ref{fig-analytical-passive}b). If we further increase $\rho$, the vesicle will be easily engulfed (Fig. \ref{fig-analytical-passive}c), but for very high $\rho$ the vesicle will not be fully engulfed rather will stop before full engulfment (Fig. \ref{fig-analytical-passive}d). The analytic phase transition line is shown in Fig. \ref{fig:passive-compare}, compared to the simulation results (Fig. 1(a), main paper). The analytic model captures correctly the qualitative features of the engulfment transition line.

\subsection{Variation of engulfment time with $\rho$ for passive case}
\begin{figure}[ht!]
\centering
\includegraphics[scale=0.7]{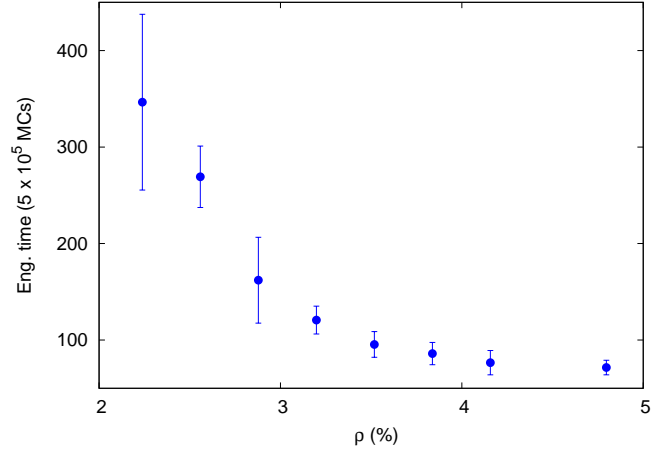}
\caption{Engulfment time with $\rho$ for passive case. The engulfment time diverges for critical $\rho \sim 2.25 ~\%$, below which there is no engulfment. Here, we use $R=10.0 l_{min}$ and $E_{ad}=1.0 ~k_BT$.}
\label{fig-eng-time-passive}
\end{figure}

We measure the engulfment time for a given $E_{ad}$ and various values of $\rho$ in Fig. \ref{fig-eng-time-passive}. We note that the engulfment time diverges as we reach the critical $\rho$ ($\sim 2.25~ \%$).

\subsection{Calculation of contributions due to active forces}
\begin{figure}[ht]
\centering
\includegraphics[scale=1]{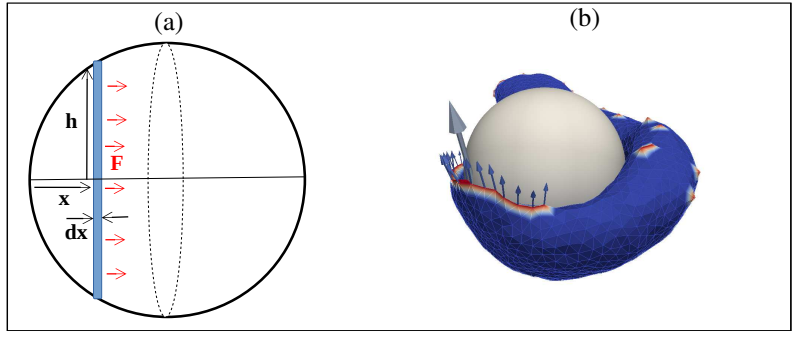}
\caption{Calculation of contribution due to active forces in the engulfment process. (a) A schematic diagram for the calculation of active work done in the engulfment process. (b) A typical snapshot of vesicle engulfing a spherical particle forming an arc-like protrusion. The arrows are showing the direction of force which is perpendicular to the radial direction of the particle. The parameters used here are $E_{ad}=1.0 ~k_BT$, $\rho=1.6 ~\%$ and $F=2.0 ~k_BT/l_{min}$.}
\label{fig:active-contribution}
\end{figure}

In this section, we estimate the contribution of active forces in the engulfment process by calculating the work done by the active forces. Let us assume a force of magnitude $F$ is acting on the membrane in the tangential direction of the surface of the particle (perpendicular direction of the radial vector of the particle) due to which the membrane extends to an infinitesimally small width $dx$ (see Fig. \ref{fig:active-contribution}a). The circumferential length of the small extended strip $dl=2 \pi h$. Thus, the work done by the active force in this process is, $dW_F = F dx $, and the increase in the adhered area $dA_{ad}=dx~ dl = 2 \pi h dx$. Thus, the active energy gain, or the active energy per unit adhered area is given by,

\begin{equation}
\partial W_F/\partial A_{ad} = F/dl = F/2 \pi h
\end{equation}

In our simulation, we choose a region close to the spherical particle (the region in between two spheres of radius $R$ and $R+R_c$, where we assume $R_c=3~l_{min}$), and assume that all the proteins inside this region is contributing in the engulfment process. We then calculate the component of force of all the proteins directed in the perpendicular direction of the radial vector of the particle, as shown by arrows in Fig. \ref{fig:active-contribution}b. The large arrow is showing the direction of net contribution. The rim length $dl$ can be estimated from the adhered fraction of area, and thereby the contribution due to active forces  can easily be estimated.

\subsection{Engulfment by RAW 264.7 macrophages }
\begin{figure}[ht!]
\centering
\includegraphics[scale=0.5]{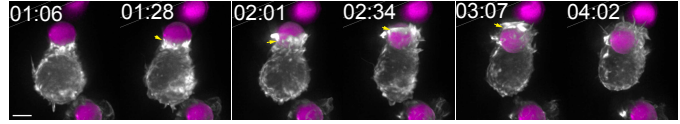}
\caption{Engulfment of Immunoglobulin G-coated polystyrene particle by RAW 264.7 macrophage. Frames from a lattice light sheet movie (maximum intensity projections) of a cell transfected with mEmerald-Lifeact are shown. Time is indicated in min:sec, scale bar $5~\mu m$}
\label{fig:experiment-si}
\end{figure}

Here, we show another dataset of the polystyrene particle engulfment by a RAW 264.7 macrophage using particle functionalized with immunoglobulin G (Fig. \ref{fig:experiment-si}).

\subsection{Configurations for large force (F)} 
\begin{figure}[ht]
\centering
\includegraphics[scale=1.4]{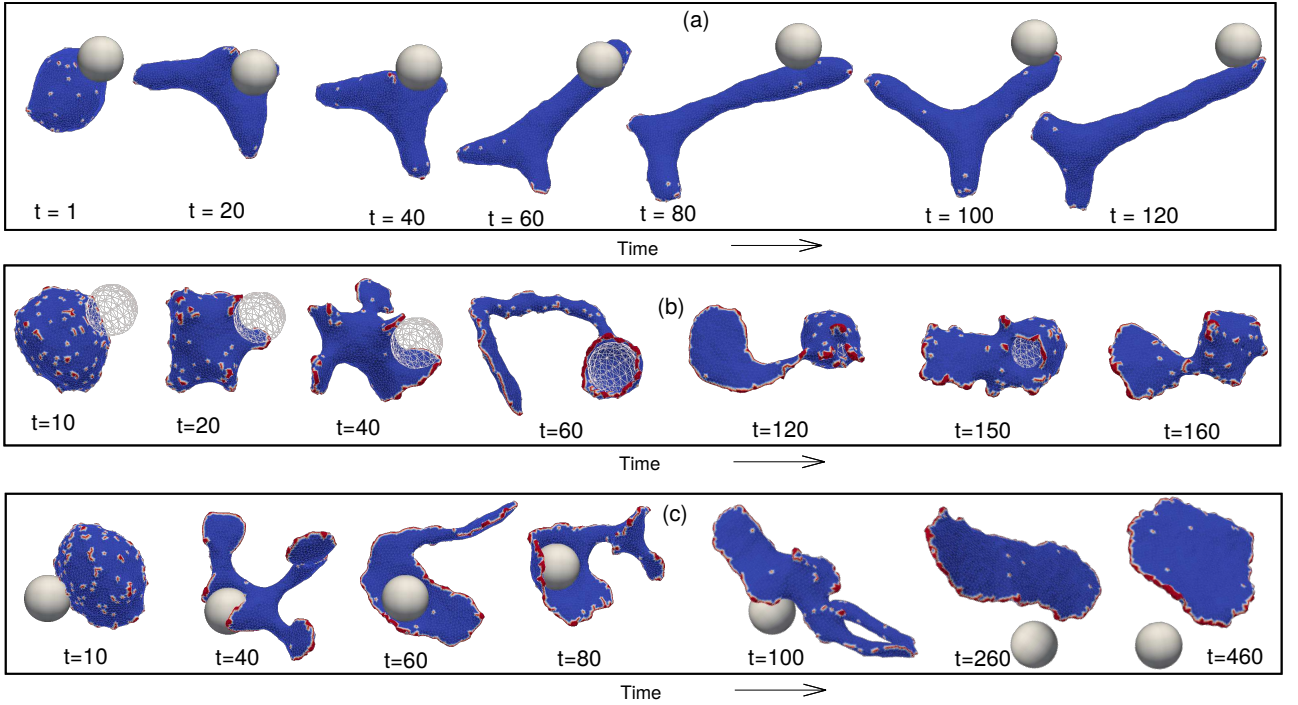}
\caption{Engulfment of spherical particles by the vesicle with large force. (a) Configuration For very large force but small protein density. The engulfment becomes difficult and the particle ends up with no engulfment in this case. Here, we use $E_{ad}=1.0 ~k_BT$, $\rho=1.6~\%$ and $F=3.0 ~k_BT/l_{min}$. (b) Configurations for a complete engulfment with very large $\rho$ and $F$. Here, we use $E_{ad}=1.0 ~k_BT$, $\rho=9.6~\%$ and $F=3.50 ~k_BT/l_{min}$. (c) Configurations for no engulfment with very large $\rho$ and even larger $F$. Here, we use $E_{ad}=1.0 ~k_BT$, $\rho=9.6~\%$ and $F=4.0 ~k_BT/l_{min}$.}
\label{large-F}
\end{figure}

For a given $E_{ad}$ and $\rho$, if we increase $F$ to very large value, at some large value of $F$, the vesicle will deadhere from the particle, and will end up with no engulfment. In Fig. \ref{large-F}a, we show such an example for $E_{ad}=1.0 ~k_BT$ and $F=3.0~ k_BT/l_{min}$ for very small protein density ($=1.6 ~\%$). We note that vesicle gets deadhered from the surface of the particle with time, and finally ends up with no engulfment (Movie-S14).

In the regime of large $F$, if we further increase $\rho$, a free vesicle (without adhesive spherical particle) can transform into a pancake-like shape \cite{miha2019}. In this regime, if we allow the vesicle to engulf the particle, it can engulf the particle if the force is not too large (Fig. \ref{large-F}b). Since the engulfment process here is faster than the pancake transition process, the vesicle finally engulfs the particle before it transform into a pancake-like shape (Movie-S15). However, if we further increase $F$ to even large value, we will again reach to a regime, where we do not have any engulfment (Fig. \ref{large-F}c, Movie-S16).

\subsection{Reorientation of vesicle with respect to non-spherical particles of high aspect ratio}
\begin{figure}[h!]
\centering
\includegraphics[scale=0.7]{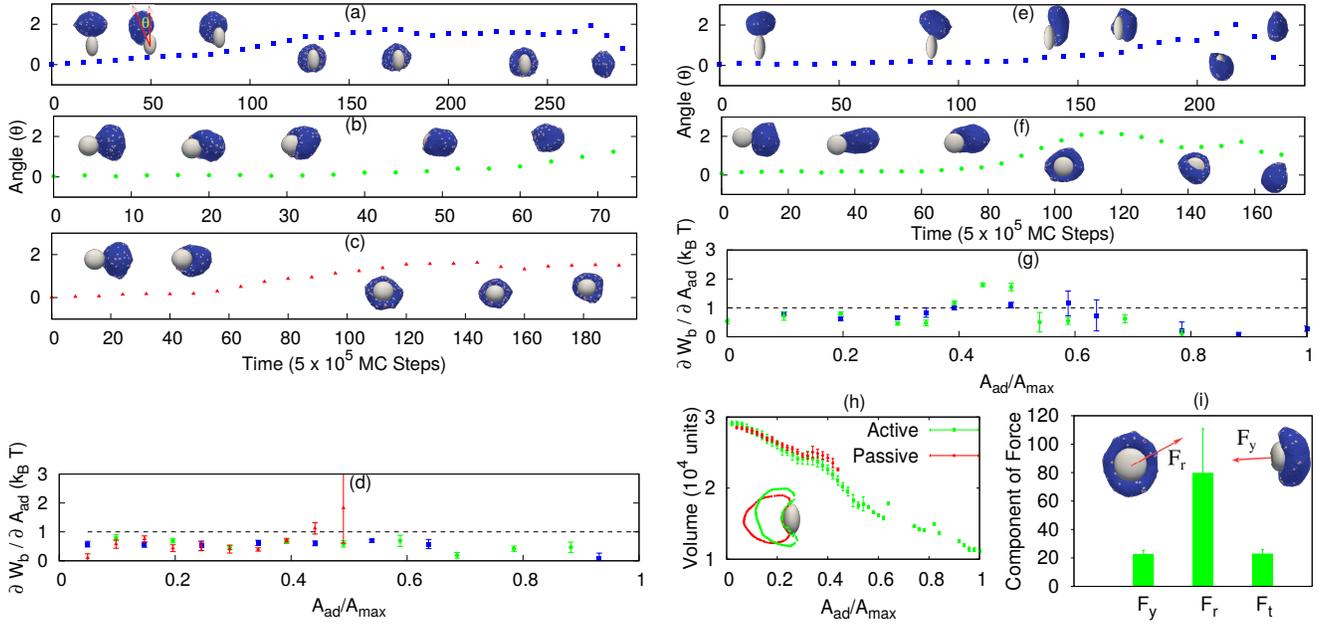}
\caption{Reorientation of the vesicle with respect to its initial position on the surface of spheroid particles of high aspect ratio. The reorientation of the vesicle is quantified in (a-c) by the angular motion of the center of mass with respect to particle's surface (denoted in the inset of (a)). (a) A vesicle with passive proteins initiated from the top of a prolate spheroid, rotates over time, and finally completes the engulfment from the side. Here, we use $R_x=R_y=8.4~ l_{min}$, $R_z=14.16~ l_{min}$. (b-c) A passive vesicle is initiated from the side of an oblate shape and fully engulfs from side in (b), while it rotates to the top of the oblate shape and stalls around half-engulfment in (c). Here, we use $R_x=R_y=11.4 ~l_{min}$, $R_z=7.96 ~l_{min}$. (d) Comparison of bending energy cost $\partial W_b/\partial A_{ad}$ and adhesion gain ($E_{ad}$, dashed line) per adhered node (both having units of $k_BT$), for (a-c). (e) A vesicle with active curved proteins is initiated at the top of a prolate shape, rotates and engulfs from side. Here, we use $R_x=R_y=7.6 ~l_{min}$, $R_z=16.29~ l_{min}$. (f) Active vesicle initiated at the side of an oblate shape finally engulfs from the top. Here, we use $R_x=R_y=12.0~ l_{min}$, $R_z=6.77 ~l_{min}$. (g) Comparison of bending energy cost and adhesion gain ($E_{ad}$, dashed line) per adhered node, for (e-f). (h) Variation of the vesicle volume (in units of $10^4~ l_{min}^3$) for both active and passive proteins, initiated from the side of an oblate shape. The inset is showing the cross section of typical snapshots for both case for $A_{ad}/A_{max} \sim 0.45$. (i) Component of the active force along different direction for the active vesicle of (h), with $A_{ad}/A_{max} \sim 0.45$. For (h-i), we use $R_x=R_y=12.0 ~l_{min}$, $R_z=6.77~ l_{min}$. For the passive case we use $\rho=4.8 ~\%$ and for the active case, $\rho=1.6~\%$, $F=1.0 ~k_BT/l_{min}$. We used $E_{ad}=1.0 ~k_BT$ for all the plots.}
\label{fig:rotation}
\end{figure}

In Fig. \ref{fig:rotation}a we show the reorientation process, first for a vesicle with passive curved proteins, that is initially in contact with a prolate particle from the top. Since, the top of a prolate shape with high aspect ratio is highly curved, the vesicle rotates and moves to the side of the particle, and finally engulfs it (Movie-S17). We quantify this rotation by the angular motion ($\theta$) of the center of mass of the vesicle with respect to the particle (see \ref{fig:rotation}a, second inset). The angle becomes close to $\pi/2$ when the vesicle rotates completely from top to side, and then it engulfs the particle. The reorientation process in this case is fast, thereby, the total engulfment times for either top or side initial conditions are very similar (blue circles and red boxes in Fig. 5, main paper, for $R_x/R_z<1$).

The reorientation of the vesicle from its initial position may not always be beneficial for engulfment. For an oblate particle with high aspect ratio,  we note that the passive vesicle can engulf the particle from side quite easily (Fig. \ref{fig:rotation}b; Movie-S18). However, we frequently observed that an initial fluctuation caused the vesicle to slide over to the flat top regions, where it maximizes its adhesion energy at low bending cost. However, in this position the vesicle has to overcome a large bending energy barrier along the entire sharp equator of the oblate shape. For $R_x=R_y>11~l_{min}$ this barrier leads to no engulfment from the top orientation, and therefore reorienting to the top from the side will end up with no engulfment (Fig. \ref{fig:rotation}c; Movie-S19) \cite{Jeroen2009}. In this case, we calculate the engulfment time by averaging only those realizations that end up with complete engulfment, while we discard the non-engulfed cases that shifted to the top position. 

In Fig. \ref{fig:rotation}d we compare the bending energy cost and adhesion energy gain ($E_{ad}$, dashed line) per adhered node, for the vesicles with passive proteins of Fig. \ref{fig:rotation}(a-c). Similar to the engulfment of spheres (Fig. 1(b), main paper), we note that for the partial engulfment the bending energy cost increases above the adhesion strength, while for the complete engulfment the bending energy cost per node remains lower than $E_{ad}$.

For vesicles with active proteins we find similar reorientation of the vesicle with respect to its initial position on the particle. In Fig. \ref{fig:rotation}e we show such a case for a prolate particle with high aspect ratio. Similar to the passive case (Fig. \ref{fig:rotation}a), the vesicle rotates to the side, to avoid the highly curved pole, and then engulfs the particle from the side (Movie-S20). Since the active case is more dynamic, the vesicle sometime takes more time to reach the side orientation, and thereby the engulfment time is larger than when it is started from side. We also show similar observation for an oblate particle with high aspect ratio, where an active vesicle rotates from the side of an oblate particle to the top position (similar to the passive case Fig. \ref{fig:rotation}c) and then engulfs it (Fig. \ref{fig:rotation}f, Movie-S21). For an oblate shape, the engulfment is easier from the side (as shown for the passive case Fig. \ref{fig:rotation}b), but since the active vesicle is very dynamic, it does not engulf from side, but rather it reorients to the top in every realization and finally engulfs. This reorientation is however very fast, as the vesicle gains large adhesion energy in this process, thereby the engulfment time starting from either top or side are comparable (green circles and magenta boxes in Fig. 5, main paper, for $R_x/R_z>1$). 

We are therefore presented with a puzzle: what allows the active proteins to robustly engulf the oblate particle from the top orientation, whereas the passive vesicle faced a bending energy barrier that stalled the engulfment. This puzzle is further emphasized when we compare the bending energy cost and adhesion energy gain ($E_{ad}$) per adhered node, for both the prolate and oblate cases in Fig. \ref{fig:rotation}g. We note that for the oblate shape the bending energy cost can increase above the adhesion strength, as for the passive system (Fig. \ref{fig:rotation}d), but nevertheless the engulfment process does not stop. In this case, the active forces provide a mechanism to engulf the particle which we describe below.

For an oblate shape with high aspect ratios, we find that the active forces do not induce the formation of an arc-like aggregate of proteins at the rim of the phgocytic cup, as happens during the engulfment of the spherical particle (Figs. 2(c-e), main paper). The active forces rather stretch the vesicle sideway, perpendicular to the particle surface, and this allows the stretched membrane afterwards to bend over the highly curved equator region of the oblate shape, leading to complete engulfment (Fig. \ref{fig:rotation}f). On the other hand, the passive vesicle is not stretched and remains spherical-like and fails to engulf the particle even for a much larger protein density (Movie-S22). This stretching of the vesicle by the active forces is quantified by measuring the volume of the vesicle, which becomes much lower than that of the passive vesicle as the adhered fraction increases (Fig. \ref{fig:rotation}h). We also show the cross-section view of both the passive and active vesicles for $A_{ad}/A_{max} \sim 0.45$ in the inset of Fig. \ref{fig:rotation}h, where the engulfment by the passive vesicle is stalled. Since the vesicle gets stretched for the active case, there is a significant bending energy contribution which arises far away from the surface of the particle, while for the passive vesicle the bending is dominated by the highly curved rim along the vesicle-particle leading edge  (Fig. \ref{oblate-mean-c}). This indicates that the bending energy cost shown in Fig. \ref{fig:rotation}g also has a significant contribution due to a membrane that is far away from the particles, and is therefore not directly affected by the progression of the membrane-particle adhesion. The sideways stretching by the active forces is demonstrated by the force components plotted in Fig. \ref{fig:rotation}i. At the time where the vesicle is poised at the highly curved equator of the oblate shape, the vesicle is getting stretched by forces that are directed radially outward ($F_r$), while the engulfment is proceeding along the $y$ direction. We note that the radial component $F_r$ is much larger that the $F_y$ component. We therefore identify here another mechanism by which active forces enable the engulfment to overcome bending energy barriers, which is qualitatively different from the mechanism described in Fig. 2, main paper. In Fig. \ref{fig:oblate-config}, we show some configurations close to the engulfment time to show how the stretching of the vesicle finally leads to full engulfment of the oblate particle.

\subsubsection{Engulfment of oblate particles: Active vesicle shows high mean curvature far away from the particle}
\begin{figure}[ht]
\centering
\includegraphics[scale=0.67]{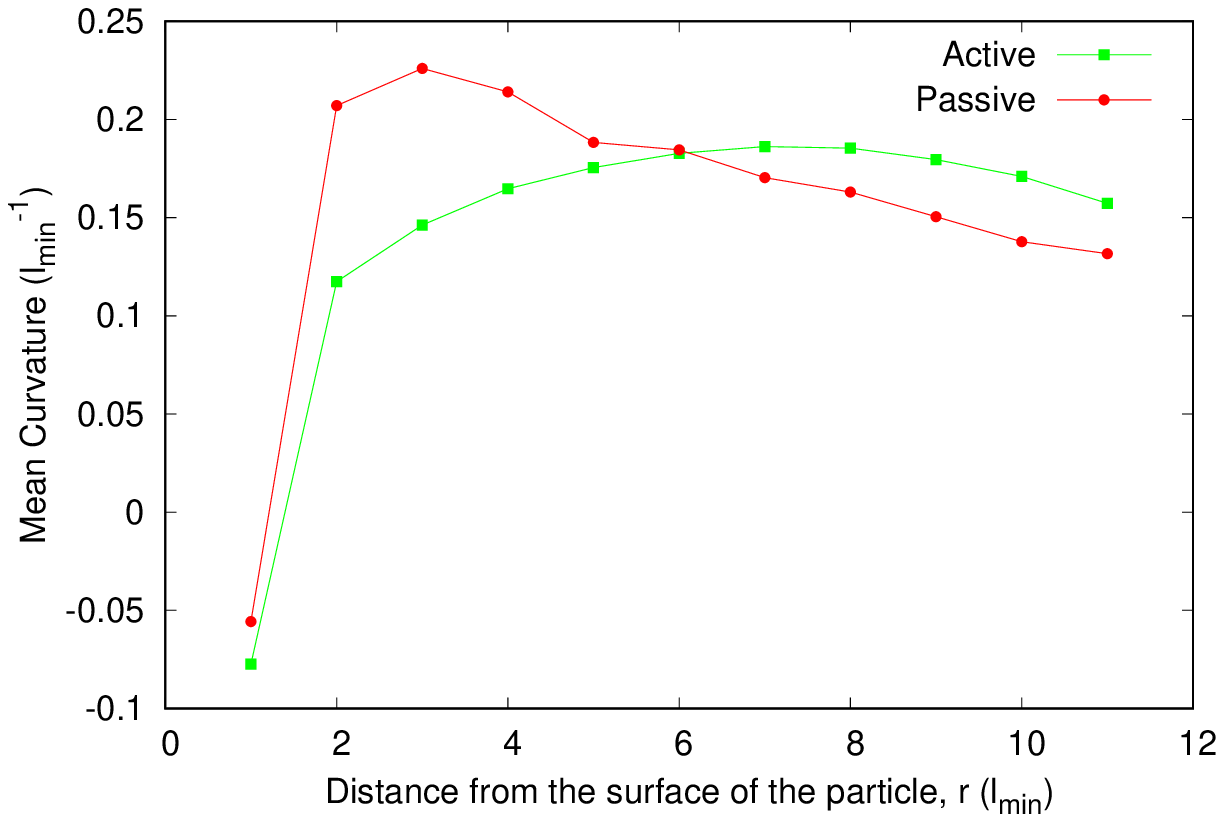}
\caption{Mean curvature at a distance $r$ from the surface of an oblate particle for $A_{ad}/A_{max} \sim 0.45$. Here, we use $R_x=R_y=12.0~ l_{min}$, $R_z=6.77 ~l_{min}$ for the oblate particle. For passive case, we use $\rho=4.8 \%$. For active case, we use $\rho=1.6~\%$ and $F=1.0~k_BT/l_{min}$. We use $E_{ad}=1.0~k_BT$ for both the cases. }
\label{oblate-mean-c}
\end{figure}

\begin{figure}[ht]
\centering
\includegraphics[scale=0.7]{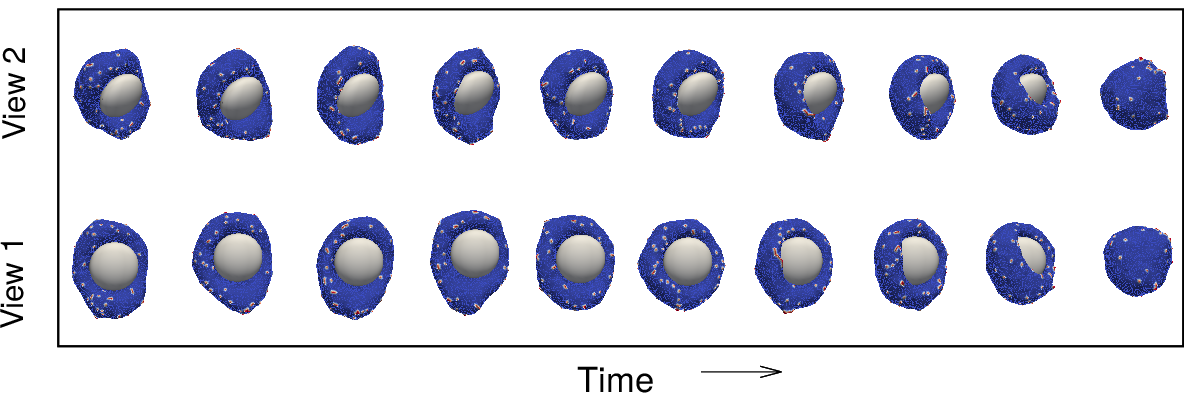}
\caption{Configurations for the engulfment of an oblate by an active vesicle close to the engulfment time.}
\label{fig:oblate-config}
\end{figure}

In Fig. \ref{fig:rotation}h, we compare the active and passive cases for the engulfment of an oblate particle with high aspect ratio, when the adhered area fraction $A_{ad}/A_{max} \sim 0.45$. Since the active vesicle gets stretched by the active forces, and thereby the bending energy cost will have dominant contribution far away from the surface of the particle. In order to quantify this, we measure the mean curvature of the vesicle section, which is having a perpendicular distance from the surface of the particle in between $r$ and $r-1$ and plot it as a function of $r$ for both the active and passive cases in Fig. \ref{oblate-mean-c}. As expected, the active vesicle have mean curvature larger than passive far away from the surface of the particle and it drops slowly with $r$. On the other hand, the passive vesicle has larger contribution only for small $r$ and it drops faster with $r$.

\subsubsection{Configurations for the engulfment of oblate shape by active vesicle close to the engulfment time}
Here, we show the configurations for the engulfment of an oblate shape by an active vesicle, discussed in Fig. \ref{fig:rotation}(h-i). Here, we zoom into the time close to the engulfment to observe how the full engulfment proceeds.

\subsection{Engulfment of bacteria-like particles}
Bacterial pathogens are often engulfed by phagocytic cells, as well as by cells which they invade \cite{Klemens2005}. We therefore investigate the engulfment of a sphero-cylindrical shape, resembling the shapes of different bacterias such as \textit{E. coli} or \textit{Bacillus subtilis} \cite{moller2012race,Richards2016}. This shape is an extreme case of a prolate spheroid, with an aspect ratio of $1:5$. We use here a larger vesicle size ($N = 6127$), since the large aspect ratio of the particle requires a larger membrane area for engulfment. 

\begin{figure}[h!]
\centering
\includegraphics[scale=0.7]{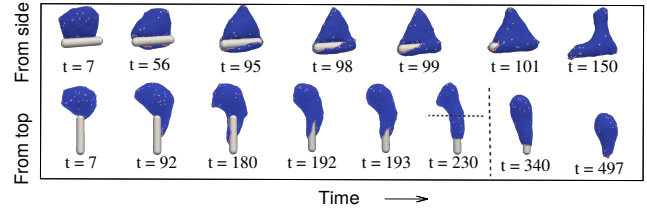}
\caption{Engulfment of a sphero-cylindrical shape. The configurations are shown as function of the simulation time for a vesicle with active curved proteins, when initiated from the side (top panel), or from the top (bottom panel). The dashed vertical line in the lower panel shows the time at which we switch off the active forces of proteins that located above the horizontal dashed line shown for $t=230$. The vesicle parameter values  are $N=6127$, $E_{ad}=1.5~k_BT$, $F=1.5~k_BT/l_{min}$, $\rho=1.6 ~\%$, while for the sphero-cylindrical shape: total length $l=60~ l_{min}$ and radius $r=6~ l_{min}$, such that the aspect ratio is $1:5$.}
\label{fig:spherocylinder}
\end{figure}

When the initial contact between the vesicle and the particle is from the side, the engulfment proceeds smoothly (Fig. \ref{fig:spherocylinder}, upper panel, Movie-S23). However, when initiated from the top, the initial engulfment stage is faster but later on the engulfment slows down and almost stalls (Fig. \ref{fig:spherocylinder}, lower panel, before the dashed vertical line). This type of stalling is also observed in experiments \cite{Richards2016}. In our simulations the stalling arises due to membrane tension, caused by active proteins that are located away from the engulfed article, and pull the membrane in the opposite the direction. We demonstrate this by turning off the active forces for all proteins that are located above the horizontal dashed line shown for $t=230$, i.e. all the proteins that are on the opposite side of the phagocytic cup are made passive. The dashed vertical line in Fig. \ref{fig:spherocylinder} lower panel shows the time at which we turn off these active proteins, and we note that as a result the engulfment process proceeds quickly afterwards to completion (Movie-S27).

We can compare these simulation results to observed bacterial phagocytosis. In \cite{moller2012race}. it was found that the velocity of engulfment was faster when the initial contact was from the pole of the bacteria. We find a similar behavior, as long as there is no restriction from membrane tension. Such tension can stall the engulfment in our simulations (Fig. \ref{fig:spherocylinder} lower panel), and indeed highly filamentous bacteria are found to avoid engulfment by immune cells \cite{prashar2013filamentous}, which we can attribute to the restriction of finite membrane area and the resulting membrane tension. The critical role of membrane tension during phagocytosis was previously shown experimentally \cite{masters2013plasma}.

\begin{figure}[ht]
\centering
\includegraphics[scale=0.7]{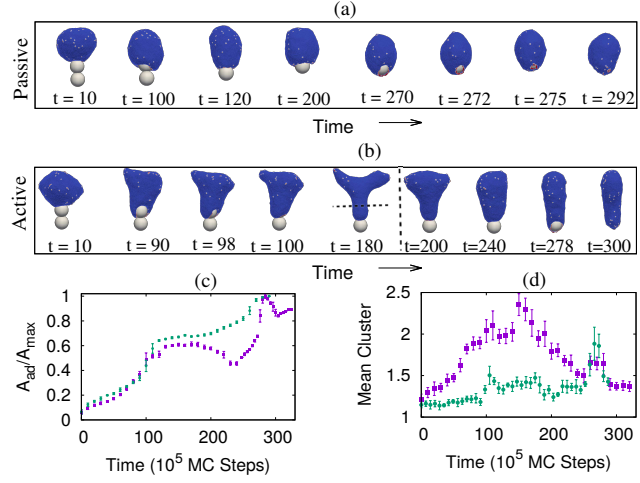}
\caption{Engulfment of dumb-bell shaped particle. (a) Configurations with time for passive case. (b) Configurations with time for active case. The vertical dashed line is showing the time, when we remove the force from the proteins at the back. The horizontal dashed line on the configuration corresponding to $t=180$ shows the coordinate above which we switch off the active forces of all the proteins. (c) Adhered area fraction with time. (d) Mean cluster size with time. Blue boxes are for active case and green circles are for passive case. The parameter values  are $N=6127$, $E_{ad}=1.5~k_BT$, $\rho=1.6 ~\%$. The radius of the spherical beads are $10~ l_{min}$ each, and the overlapped region $l_p=2~ l_{min}$, such that total length $L=4 R - l_p=38~l_{min}$. For active case, we use $F=1.5~k_BT/l_{min}$. }
\label{Dumb-bell}
\end{figure}

We also explore the engulfment of dumb-bell like shapes, resembling the shape of fungi such as budding yeast \cite{Clarke2010}. A similar effect of membrane tension is also observed in our simulations during the engulfment such dumb-bell shaped particle (Fig. \ref{Dumb-bell}, Movie-S25,S26). We found that while passive curved proteins can give rise to a smooth engulfment that does not stall (Fig. \ref{Dumb-bell}a), the active vesicle can easily stall after engulfment of one lobe of the particle (i.e. when the leading edge is at the narrow neck)  (Fig. \ref{Dumb-bell}b). Releasing the membrane tension induced by the active proteins at the back of the vesicle, can release the vesicle and allow full engulfment. Similar hourglass shapes were previously shown theoretically to be susceptible to stalled engulfment \cite{Richards2016}, as well as in experiments observing phagocytosis of dividing yeast cells \cite{Clarke2010,dieckmann2010myosin} or highly deformable particles \cite{Nils2021}. We also show the mean cluster size and the adhered fraction for both the passive and the active case in Fig. \ref{Dumb-bell}(c-d). 

\subsection{Results for multicomponent membrane with proteins of different intrinsic curvature}
\begin{figure}[ht!]
\centering
\includegraphics[scale=1.4]{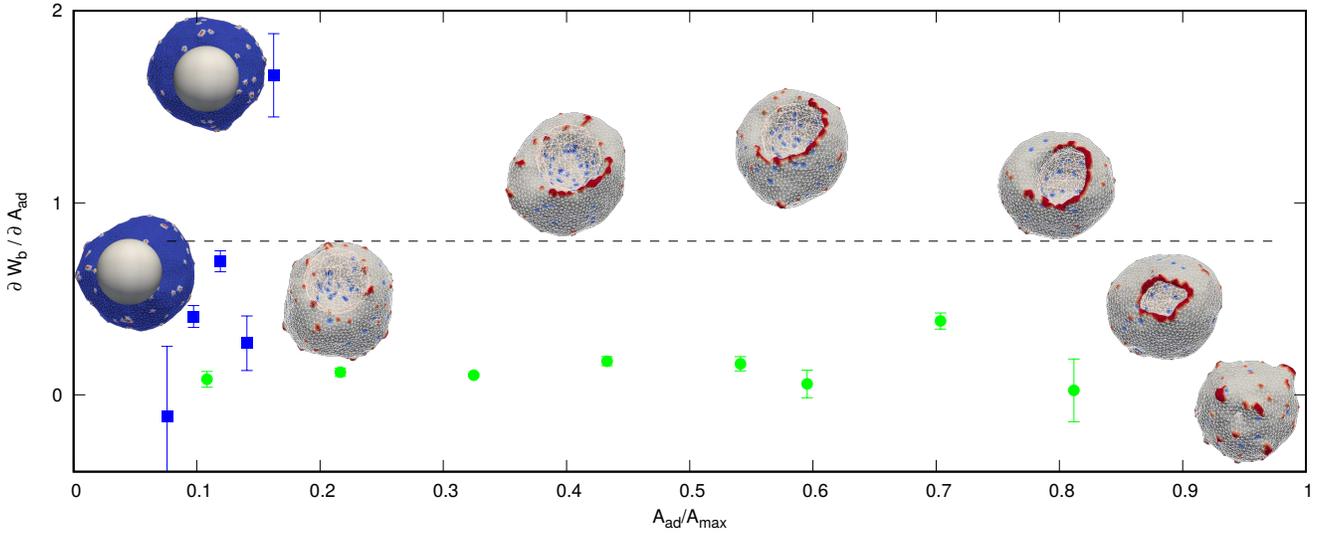}
\caption{Results for multicomponent membrane with proteins of different intrinsic curvature. Here, we show the bending energy cost with adhered area fraction, and compare it with the adhesion strength $E_{ad}$. Blue boxes are for convex proteins, and green circles are for multiple proteins. Left insets are for convex proteins and right insets are for multiple curvature proteins. For multiple curvature case, red regions are denoting convex proteins, and blue regions are denoted concave proteins. Here, we use $R=10.0~l_{min}$, $E_{ad}=0.80 ~k_BT$, and $\rho_p=3.2~\%$ for convex protein case. For multiple proteins case, we have in addition $\rho_n=3.2~\%$. }
\label{fig:multiple}
\end{figure}

Here, we show our results for multicomponent membrane with proteins of different intrinsic curvature. We consider convex ($c_0=1.0~ l_{min}^{-1}$) as well as concave ($c_0=-1.0 ~l_{min}^{-1}$) proteins. We note that addition of concave proteins makes the engulfment easier in the parameter regime where only convex proteins fails to engulf the particle. In Fig. \ref{fig:multiple}, we show the engulfment by passive proteins with small  $E_{ad}$ ($=0.80 ~k_BT$) and convex protein density $\rho_p=3.2~\%$ such that there is no engulfment, and compare with the case where in addition, we have concave proteins of density $\rho_n=3.2~\%$. The convex (concave) proteins have attractive interaction only with the convex (concave) one. We note that the concave proteins aggregate close to the particle, and reduces the bending energy there. This allows the vesicle to adhere more and engulf the particle (Movie-S24).


\section{Supplementary movies}
All the supplementary movies are available online in the link: https://app.box.com/s/ske4vjlp6ci3vu7mz5izfp2iebxrmgm4 \\

\begin{itemize}

\item{\textbf{Movie-S1 }Complete engulfment of a spherical particle of larger radius by a protein-free vesicle. Here, we use $R=10~ l_{min}$ and $E_{ad}=1.30~ k_BT$.}
\label{movie1}

\item{\textbf{Movie-S2 }Partial engulfment of a spherical particle of smaller radius by a protein-free vesicle. We use here $R=8 ~l_{min}$ and $E_{ad}=1.30~ k_BT$.}
\label{movie2}

\item{\textbf{Movie-S3 }Partial engulfment of a spherical particle by a passive vesicle with small protein density. We use $R=10 ~l_{min}$, $\rho=2.4~\%$ and $E_{ad}=1.0 ~k_BT$.}
\label{movie3}

\item{\textbf{Movie-S4 }Complete engulfment of a spherical particle by a passive vesicle with medium protein density. Here, we use $R=10 ~l_{min}$, $\rho=4.0~\%$ and $E_{ad}=1.0~ k_BT$.}
\label{movie4}

\item{\textbf{Movie-S5 }Partial engulfment of a spherical particle by a passive vesicle with very high protein density. Here, we sue $R=10 ~l_{min}$, $\rho=19.2\%$ and $E_{ad}=1.0~ k_BT$.}
\label{movie5}

\item{\textbf{Movie-S6 }Partial engulfment of a spherical particle by an active vesicle with small force. The parameters are: $R=10 ~l_{min}$, $\rho=1.6~\%$, $E_{ad}=1.0~ k_BT$ and $F=0.10 ~k_BT/l_{min}$.}
\label{movie6}

\item{\textbf{Movie-S7 }Complete engulfment of a spherical particle by an active vesicle with medium force. Here, we use $R=10 ~l_{min}$, $\rho=1.6~\%$, $E_{ad}=1.0 ~k_BT$ and $F=0.40 ~k_BT/l_{min}$.}
\label{movie7}

\item{\textbf{Movie-S8 }Complete engulfment of a spherical particle by an active vesicle with large force. We use here $R=10 ~l_{min}$, $\rho=1.6~\%$, $E_{ad}=1.0 ~k_BT$ and $F=2.0~ k_BT/l_{min}$.}
\label{movie8}

\item{\textbf{Movie-S9 }Engulfment of spherical particle by an active vesicle with large $\rho$ and small $F$. Here, we show fragmented protein clusters on the phagocytic cup at the early stage of engulfment. We use here $R=10 ~l_{min}$, $E_{ad}=1.5 ~k_BT$, $\rho=6.4 ~\%$ and $F=1.0~ k_BT/l_{min}$. }
\label{movie9}

\item{\textbf{Movie-S10 }Engulfment of spherical particle by an active vesicle with large $\rho$ and large $F$. Here, we use $R=10 ~l_{min}$, $E_{ad}=1.5 ~k_BT$, $\rho=4.8 ~\%$ and $F=2.0~ k_BT/l_{min}$. }
\label{movie10}

\item{\textbf{Movie-S11 }Phagocytosis of Immunoglobulin-G-coated polystyrene bead (blue) by RAW264.7 macrophage transfected with mEmerald-Lifeact (orange). Maximum intensity projection of lattice light-sheet imaging. Scale bar, $5~\mu m$. Time stamp: min:sec.}
\label{movie11}

\item{\textbf{Movie-S12 }Phagocytosis of Immunoglobulin-G-coated polystyrene bead (maroon) by murine bone-marrow derived macrophage transfected with mEmerald-Lifeact (gray). Maximum intensity projection of lattice light-sheet imaging. Scale bar, $5~\mu m$. Time stamp: min:sec.}
\label{movie12}

\item{\textbf{Movie-S13 }Phagocytosis of Immunoglobulin-G-coated polystyrene bead (maroon) by RAW264.7 macrophage transfected with mEmerald-Lifeact (gray). Maximum intensity projection of lattice light-sheet imaging. Scale bar, $5~\mu m$. Time stamp: min:sec.}
\label{movie13}

\item{\textbf{Movie-S14 }No engulfment of a spherical particle by an active vesicle with very large force. The movie shows a case, where particle is not engulfed by the active vesicle. Here, we use $R=10~l_{min}$, $\rho=1.6~\%$, $E_{ad}=1.0 ~k_BT$ and $F=3.0~ k_BT/l_{min}$.}
\label{movie14}

\item{\textbf{Movie-S15 }Complete engulfment of a spherical particle by an active vesicle with very large force and large protein density. Here, we use $R=10~l_{min}$, $\rho=9.6~\%$, $E_{ad}=1.0 ~k_BT$ and $F=3.50~ k_BT/l_{min}$.}
\label{movie15}

\item{\textbf{Movie-S16 }No engulfment of a spherical particle by an active vesicle with even larger force and large protein density. We use here $R=10~l_{min}$, $\rho=9.6~\%$, $E_{ad}=1.0 ~k_BT$ and $F=4.0 ~k_BT/l_{min}$.}
\label{movie16}

\item{\textbf{Movie-S17 }Complete engulfment of a prolate spheroid of high aspect ratio by a passive vesicle. The vesicle is initially started from the top and then rotates to the side and engulfs the particle. Here, we use $R_x=R_y=8.4~ l_{min}$, $R_z=14.16 ~l_{min}$, $\rho=4.8~\%$ and $E_{ad}=1.0~ k_BT$.}
\label{movie17}

\item{\textbf{Movie-S18 }Complete engulfment of an oblate spheroid of high aspect ratio by a passive vesicle. The vesicle is initially started from the side, it engulfs the particle without reorienting itself to the other side. Here, we use $R_x=R_y=11.4 ~l_{min}$, $R_z=7.96 ~l_{min}$, $\rho=4.8~\%$ and $E_{ad}=1.0 ~k_BT$.} 
\label{movie18}

\item{\textbf{Movie-S19 }Partial engulfment of an oblate spheroid of high aspect ratio by a passive vesicle. The vesicle is initially started from the side, and it rotates to the top and ends up with no engulfment. The parameters are same as the previous movie.}
\label{movie19}

\item{\textbf{Movie-S20 }Complete engulfment of a prolate spheroid of high aspect ratio by an active vesicle. The vesicle is initially started from the top, and then reorients itself to the side and engulfs the particle. Here, we use $R_x=R_y=8.4 ~l_{min}$, $R_z=14.16 ~l_{min}$, $\rho=1.6~\%$, $E_{ad}=1.0 ~k_BT$ and $F=1.0 ~k_BT/l_{min}$.}
\label{movie20}

\item{\textbf{Movie-S21 }Complete engulfment of an oblate spheroid of high aspect ratio by an active vesicle. The vesicle is initially started from the side, and rotates to the top and engulfs the particle. Here, we use $R_x=R_y=12.0 ~l_{min}$, $R_z=6.77~ l_{min}$, $\rho=1.6~\%$, $E_{ad}=1.0 ~k_BT$ and $F=1.0 ~k_BT/l_{min}$.}
\label{movie21}

\item{\textbf{Movie-S22 }Comparison of active and passive engulfment of an oblate shaped particle with high aspect ratio, initiated from side. For passive case the engulfment is partial while for active case, the engulfment is complete. Here, we use $R_x=R_y=12.0 ~l_{min}$, $R_z=6.77~ l_{min}$, $E_{ad}=1.0 ~k_BT$. For active case we use $\rho=1.6 ~\%$ and $F=1.0 ~k_BT/l_{min}$. For passive case, we use $\rho=4.8 ~\%$ }
\label{movie22}

\item{\textbf{Movie-S23 }Engulfment of a sphero-cylinder by an active vesicle, when initiated from the top (from the spherical cap). Here, we use total number of vertices on the vesicle, $N=6127$ radius of spherical cap $r=6 ~l_{min}$, total length $l=60 ~l_{min}$, $\rho=1.6~\%$, $E_{ad}=1.50 ~k_BT$ and $F=1.50 ~k_BT/l_{min}$.}
\label{movie23}

\item{\textbf{Movie-S24 }Engulfment of a sphero-cylinder by an active vesicle, when initiated from the side (from the cylindrical part). The parameters are same as the previous case.}
\label{movie24}

\item{\textbf{Movie-S25 }Engulfment of a dumb-bell shaped particle by a passive vesicle. Here, we use $N=6127$ radius of spherical beads $R=10 ~l_{min}$, thickness of the overlapped region of the two beads $l_p=2 ~l_{min}$, $\rho=1.6~\%$, $E_{ad}=1.50~ k_BT$.}
\label{movie25}

\item{\textbf{Movie-S26 }Engulfment of a dumb-bell shaped particle by an active vesicle. The movie shows the engulfment of a dumb-bell shaped particle by an active vesicle with $F=1.50 ~k_BT/l_{min}$. Other parameters are same as the previous case.}
\label{movie26}

\item{\textbf{Movie-S27 }Engulfment of a spherical particle by a passive vesicle with multiple curvature proteins. The movie shows a complete engulfment of spherical particle of radius $R=10.0 ~l_{min}$ by a passive vesicle, with $E_{ad}=1.0~ k_BT$, convex protein density $\rho_p=3.2~ \%$ and concave protein density $\rho_n=3.2 ~\%$. }
\label{movie27}

\end{itemize}


\end{document}